\documentclass[iop]{emulateapj}
\usepackage{natbib}
\usepackage{graphicx}
\usepackage{longtable}
\usepackage{layout}
\usepackage[british]{babel}
\usepackage[latin1]{inputenc}
\usepackage{amssymb}
\usepackage{natbib}
\usepackage{textcomp}

\slugcomment{Draft 06-2014}

\shortauthors{De Rosa et al. 2014}

\def\Civ{C\,{\sc iv}}

\def\Mgii{Mg\,{\sc ii}}
\def\Feii{Fe\,{\sc ii}}
\def\Feiii{Fe\,{\sc iii}}

\def\Lya{Ly$\alpha$}

\def\CIIforb{[C\,{\sc ii}]}
\def\CIII{C\,{\sc iii}]}
\def\SiIV{Si\,{\sc iv}}

\def\Hb{H$\beta$}

\def\lsim{\mathrel{\rlap{\lower 3pt \hbox{$\sim$}} \raise 2.0pt \hbox{$<$}}}
\def\gsim{\mathrel{\rlap{\lower 3pt \hbox{$\sim$}} \raise 2.0pt \hbox{$>$}}}

\def\Msun{M$_\odot$}

\def\Mbh{$M_{\rm BH}$}

\begin{document}


\title{Black hole mass estimates and emission-line properties of
  a sample of redshift $z>6.5$ quasars\altaffilmark{1}}

\author{Gisella~De~Rosa\altaffilmark{2,3}, Bram~P.~Venemans\altaffilmark{4}, Roberto~Decarli\altaffilmark{4}, 
Mario~Gennaro\altaffilmark{5}, Robert~A.~Simcoe\altaffilmark{6}, Matthias~Dietrich\altaffilmark{7}, Bradley~M.~Peterson\altaffilmark{2,3}, 
Fabian~Walter\altaffilmark{4}, Stephan~Frank\altaffilmark{2}, Richard~G.~McMahon\altaffilmark{8,9}, Paul~C.~Hewett\altaffilmark{8}, 
Daniel~J.~Mortlock\altaffilmark{10,11}, Chris~Simpson\altaffilmark{12}}

\altaffiltext{1}{Based on observations collected at the European Southern
Observatory, Chile, programs 286.A-5025, 087.A-0890 and 088.A-0897. This
paper also includes data gathered with the 6.5 m Magellan Telescope
located at Las Campanas Observatory, Chile.}
\altaffiltext{2}{Department of Astronomy, The Ohio State University, 140 West 18th Avenue, Columbus, OH 43210, USA}
\altaffiltext{3}{Center for Cosmology \& AstroParticle Physics, The Ohio State University, 191 West Woodruff Ave, Columbus, OH 43210, USA}
\altaffiltext{4}{Max-Planck-Institut f\"{u}r Astronomie, K\"{o}nigstuhl 17, D-69117, Heidelberg, Germany}
\altaffiltext{5}{Space Telescope Science Institute, 3700 San Martin Drive,  Baltimore, MD 21218, USA}
\altaffiltext{6}{MIT-Kavli Center for Astrophysics and Space Research, 77 Massachusetts Avenue, Cambridge, MA, 02139, USA}
\altaffiltext{7}{Department of Physics \& Astronomy, Ohio University, Clippinger Lab 251B, Athens, OH 45701, USA} 
\altaffiltext{8}{Institute of Astronomy, University of Cambridge, Madingley Road, Cambridge CB3 0HA, U.K.} 
\altaffiltext{9}{Kavli Institute for Cosmology, University of Cambridge, Madingley Road, Cambridge CB3 0HA, U.K.} 
\altaffiltext{10}{Astrophysics Group, Imperial College London, Blackett Laboratory, Prince Consort Road, London, SW7 2AZ, U.K.} 
\altaffiltext{11}{Department of Mathematics, Imperial College London, London, SW7 2AZ, U.K.} 
\altaffiltext{12}{Astrophysics Research Institute, Liverpool John Moores University, Liverpool Science Park, 146 Brownlow Hill, Liverpool L3 5RF, U.K.} 

\begin{abstract}
  We present the analysis of optical and near-infrared spectra of the only
  four $z>6.5$ quasars known to date, discovered in the UKIDSS-LAS and
  VISTA-VIKING surveys.  Our data-set consists of new VLT/X-Shooter and
  Magellan/FIRE observations. These are the best 
  optical/NIR spectroscopic data that are likely
  to be obtained for the $z>6.5$ sample using current $6$ - $10$ m facilities.
  We estimate the black hole mass, the Eddington ratio, and the \SiIV/\Civ,
  \CIII/\Civ, and \Feii/\Mgii \ emission-line flux ratios.  We perform
  spectral modeling using a procedure that allows us to derive a probability
  distribution for the continuum components and to obtain the quasar
  properties weighted upon the underlying distribution of continuum models.
  The $z>6.5$ quasars show the same emission properties as their counterparts
  at lower redshifts.  The $z>6.5$ quasars host black holes with masses of
  $\sim 10^9$ \Msun \ that are accreting close to the Eddington luminosity
  ($\langle{\rm log} (L_{\rm Bol}/L_{\rm Edd})\rangle= -0.4\pm0.2$), in agreement with
  what has been observed for a sample of $4.0<z<6.5$ quasars.  By comparing
  the \SiIV/\Civ \ and \CIII/\Civ \ flux ratios with the results obtained from
  luminosity-matched samples at $z\sim6$ and $2\leq z\leq4.5$, we find no
  evidence of evolution of the line ratios with cosmic time. We compare the
  measured \Feii/\Mgii \ flux ratios with those obtained for a sample of
  $4.0<z<6.4$ sources. The two samples are analyzed using a consistent
  procedure. There is no evidence that the  \Feii/\Mgii \ flux ratio evolves between 
  $z=7$ and $z=4$. Under the assumption that the \Feii/\Mgii \
  traces the Fe/Mg abundance ratio, this implies the presence of major
  episodes of chemical enrichment in the quasar hosts in the first $\sim0.8$
  Gyr after the Big Bang.

\subjectheadings{cosmology: observations -- quasars: general, emission lines -- galaxies: active, high redshift, formation}

\end{abstract}

\section{Introduction}
\label{sec_intro}
In the past 10 years, more than sixty quasars at $5.7 \lesssim z \lesssim 6.4$
have been discovered \citep[e.g.,][]{Fan2006,Jiang2008,Jiang2009,Venemans2007,
  Willott2007, Willott2010a,Morganson2012,Banados2013}, mainly thanks to
optical surveys such as the Sloan Digital Sky Survey
\citep[SDSS;][]{York2000}, the Canada France High-z Quasar Survey
\citep[CFHQS;][]{Willott2007}, and the Panoramic Survey Telescope \& Rapid
Response System Survey 1 \citep[Pan-STARRS1;][]{Kaiser2010}. 
{Only four quasars are known to date at $z>6.5$: ULAS
J$112001.48$+$064124.3$ (hereafter J$1120$+$0641$) at $z=7.1$, discovered by 
\citet{Mortlock2011} in the Unite Kingdom Infrared Deep Sky Survey (UKIDSS) Large Area Survey
\citep[LAS;][]{Lawrence2007}; and VIKING
J$234833.34$--$305410.0$ (hereafter J$2348$--$3054$) at $z=6.9$, VIKING
J$010953.13$--$304726.3$ (hereafter J$0109$--$3040$) at $z=6.7$, and VIKING
J$030516.92$--$315056.0$ (hereafter J$0305$--$3150$) at $z=6.6$, recently
discovered by \citet{Venemans2013} in the Visible
and Infrared Survey Telescope for Astronomy (VISTA) Kilo-Degree Infrared
Galaxy \citep[VIKING;][]{Arnaboldi2007} survey.} These very
high redshift quasars are direct probes of the Universe less than 1 Gyr after
the Big Bang.  They are fundamental in studying the physical conditions of the
Universe during the epoch of reionization
\citep[e.g.,][]{Bolton2011,Mortlock2011,Simcoe2012}, the formation and early
growth of supermassive black holes \citep[e.g.,][]{Volonteri2010,Latif2013},
the galaxy formation processes \citep[e.g.,][]{Walter2009,Wang2013}, and the
interstellar medium chemical evolution
\citep[e.g.,][]{Jiang2007,Simcoe2011,Simcoe2012}.

{ The bright emission lines in the rest-frame UV spectrum of quasars 
($\lambda_{\rm rest}\sim 1000$ - $3000$ \AA) provide insights on the properties
of the black hole (BH) and of the circumnuclear gas. For example, one can use the Doppler 
emission-line widths to estimate the mass of the BH (\Mbh) via empirical 
mass-scaling relations \citep[e.g.,][]{Vestergaard2006,Vestergaard2009}.}  
At the same time, photoionization models show that various emission-line flux 
ratios (e.g., N\,{\sc iii}]/O\,{\sc iii}], N\,{\sc v}/(C\,{\sc iv}+O\,{\sc iv}), 
and N\,{\sc v}/He\,{\sc ii}) can be used to derive chemical abundances of the 
broad line region (BLR) gas \citep[e.g.,][]{Hamann2002, Nagao2006} and, thus, to 
set constraints on the star formation history of the quasar host galaxy. In
particular, the abundance of Fe vs $\alpha$ elements (e.g., Mg and O that are
produced via $\alpha$ processes) represents a key factor in understanding the
chemical evolution of galaxies in the early Universe. { According to chemical 
evolution scenarios, the dominant source of iron (Fe) is the explosion of type Ia
supernovae (SNe Ia), that are thought to originate from intermediate-mass
stars in close binary systems and are characterized by long life-times 
\citep[$\sim 1$ Gyr after the onset of star formation, e.g., ][]{Tinsley1979,Matteucci1986}. 
On the other hand, $\alpha$ elements are assumed to be mainly produced by core collapse
supernovae (SNe of types II, Ib, and Ic), that originate from more massive
stars which explode shortly after the initial starburst ($\sim8$ Myr). The amount of Fe 
returned to the interstellar medium through SNII ejecta is rather low \citep{Yoshi1996}.
Therefore, the Fe/$\alpha$ ratio is expected to be a strong function of age in young systems, 
with the Fe enrichment being typically delayed of $\sim 1$ Gyr. However this delay can be much shorter 
($\sim 0.3$ Gyr) for massive elliptical galaxies \citep[e.g.,][]{Matteucci2003,Pipino2011}.}

{ Numerous spectroscopic studies of high redshift quasars 
\citep[e.g.,][]{Jiang2007, Kurk2007, Kurk2009, DeRosa2011,Willott2010a} have 
shown that high redshift quasars host BHs with \Mbh$\sim 10^9 \
M_\odot$} that are accreting close to the Eddington limit.  In particular, from
the consistent analysis of a sample of 22 sources with $4.0<z<6.5$,
\citet{DeRosa2011} found that, at a given luminosity, the $z>4$ sources are
accreting faster than those at low redshift \citep[see
also][]{Trakhtenbrot2011}. The average Eddington ratio for the high redshift
quasar population is { $\langle \log(L_{\rm Bol}/L_{\rm
  Edd})\rangle=-0.35\pm0.25$, while for a luminosity-matched sample at
$0.35<z<2.25$ the average Eddington ratio is $\langle \log(L_{\rm Bol}/L_{\rm
  Edd})\rangle=-0.80\pm0.24$}. At the same time, \citet{Mortlock2011} estimated 
a black hole mass 
of \Mbh$=2.0^{+1.5}_{-0.7} \times 10^9 \ M_\odot$ for J$1120$+$0641$, and a
corresponding Eddington ratio of $L_{\rm Bol}/L_{\rm
  Edd}\sim1.2^{+0.6}_{-0.5}$. The presence of quasars hosting BHs with masses
$\gtrsim 10^9 \ M_\odot$ when the Universe is less than 1 Gyr old 
challenges models of BH seed formation \citep[e.g.,][]{Volonteri2010,Latif2013}.

At low redshift, elemental abundances estimated from both emission and
intrinsic absorption lines show that quasar environments are characterized by
solar or super-solar metallicities. \citet{Jiang2007} estimated the { BLR 
metallicity} for a sample of six luminous quasars with $5.8<z<6.3$
and found super solar metallicities (typical value of $\sim$$4 Z_\odot$) and
no strong evolution in metallicity up to $z\sim6$.  The observational proxy
that is usually adopted to trace the Fe/Mg abundance ratio is the \Feii/\Mgii
\ line ratio. For $z>5.7$ quasars the \Mgii \ emission line and the strong
\Feii \ complexes are redshifted in the NIR. Previous NIR-spectroscopy works
that studied the \Feii/\Mgii \ line ratio in samples including high-$z$ quasars
\citep[e.g.,][]{Barth2003, Dietrich2003, Iwamuro2004, Jiang2007, Kurk2007}
showed an increase in the scatter of the measured \Feii/\Mgii \ line ratios at
$z\sim6$.  However, by performing a consistent analysis of a sample of 22
qgeasars with $4.0<z<6.5$, that included many of the sources analyzed in
previous studies, \citet{DeRosa2011} found no sign of evolution in the \Feii/\Mgii \ line
ratio for $4.0<z<6.5$, suggesting an early chemical enrichment of the
circumnuclear gas.

The goal of this paper is to characterize the BH masses and the emission
properties of the only four quasars known to date at $z>6.5$.  We present new
spectroscopic data for J$1120$+$0641$; observations were carried with the
X-Shooter spectrograph \citep{Vernet2011} mounted on the Very Large Telescope
(VLT). Together with the new data, we analyze all the observations of the
$z>6.5$ sources collected by our group using the VLT/X-Shooter spectrograph
and the Folded-port InfraRed Echellette spectrograph
\citep[FIRE,][]{Simcoe2013} mounted on the Magellan Telescope. The paper is
structured as follows. In Section 2 we describe the observations and the data
reduction. In Section 3 we discuss our spectral decomposition and the modeling
procedure, while in Section 4 we present the results and discuss their
implications. Finally, we give a brief summary in Section 5. We assume the
following $\Lambda$CDM cosmology throughout the paper: $H_0=70 \ $km$ \
$s$^{-1} \ $Mpc$^{-1}$, $\Omega_M=0.28$, and $\Omega_\Lambda =0.72$
\citep{Komatsu2011}.

\section{Data}
Our sample consists of the only four quasars known to date with $z>6.5$:
J$1120$+$0641$, J$2348$--$3054$, J$0109$--$3047$ and J$0305$--$3150$.  Their
absolute magnitudes range between $-26.6<$ M$_{1450,\rm{AB}}<-25.6$ (see
Table~\ref{Tab_data} for details).

\subsection{Observations and Data Reduction}
We present new X-Shooter data for J$1120$+$0641$: observations were carried
out between 2011 March and May (total exposure time 18,000 s). X-Shooter is a
medium resolution Echellette spectrograph mounted on the Cassegrain focus of
the 8.2 m VLT Kuyen telescope (UT2).  X-Shooter covers 3 wavelength regions
with 3 different spectrographs: UVB arm 3000 - 5595 \AA, VIS arm 5595 - 10240
\AA, and NIR arm 10240 - 24800 \AA.  Given the high redshift of the source,
the rest-frame UV emission lines are redshifted at $\lambda_{\rm
  obs}\gtrsim10000$ \AA. At the same time the Gunn-Peterson absorption
\citep{Gunn1965} of the Ly continuum entirely suppresses the flux at
wavelengths $\lambda_{\rm obs}< 912 \ (1+z)$ \AA \ $\sim 7500$ \AA.  Therefore
we focus only on the VIS and NIR parts of the spectrum.  We used the
0.9\arcsec$\times$11\arcsec \ slit for both VIS and NIR observations (pixel
scales equal to 0.16\arcsec \ pixel$^{-1}$ and 0.21\arcsec \ pixel$^{-1}$,
respectively), while the DIMM seeing varied between 0.6\arcsec \ and
1.7\arcsec.  The resulting spectra have a resolution $R=8800$ in the VIS bands
and $R=5300$ in the NIR.  The raw two-dimensional spectra were rectified and wavelength
calibrated by using the X-Shooter pipeline version 1.3.7
\citep{Modigliani2010}.  Extraction and flux calibration using spectrophotometric standard stars 
were instead performed by using customized IDL routines. { The final absolute flux
calibration was obtained by matching the one-dimensional spectra to the observed UKIDSS
infrared flux. UKIDSS data were collected between November 2010 and January 2011}. 
Our X-Shooter pipeline produces 4 separated spectral
segments covering 8500$-$10200 \AA, 10000$-$14200 \AA, 14000$-$18200 \AA, and
18000$-$24000 \AA.  After properly degrading the orders with higher resolution
(bluer orders) to the lower resolution of the redder orders, individual
segments were merged together by computing an inverse variance weighted
average in the overlapping regions.

Together with the new data, we analyzed all the observations of the $z>6.5$
sources that our group collected during 2011 and 2012.  X-Shooter observations
of J$2348$--$3054$ (total exposure time 8738 s) and of J$0109$--$3047$ (total
exposure time 21,600 s) were carried out in 2011 August and November. The
resolution of the spectra varies between $R=5400$ - $8800$ in the VIS bands
and between $R=4000$ - $5000$ in the NIR ones.  A detailed description of
observing settings and data reduction can be found in the discovery paper
\citep{Venemans2013}.  Observations of J$0305$--$3150$ (total exposure time of
26,400 s) and J$1120$+$0641$ (total exposure time of 54,036 s) were carried out
with the FIRE spectrograph mounted on the Magellan/Baade 6.5 m telescope. The
data were taken with a 0.6\arcsec \ slit, resulting in a spectral resolution
of $R=6000$ over the full 8200 \AA \ - 25100 \AA \ wavelength range. A
detailed description of observing conditions and data reduction can be found
in \citet{Venemans2013} for J$0305$--$3150$ and in \citet{Simcoe2012} for
J$1120$+$0641$.

\begin{deluxetable*}{lcccccc}
\tablecaption{\label{Tab_data} Quasars in current sample.}
\tablecolumns{7}
\tablewidth{0pc}
\tablehead{
\colhead{Quasar name} & \colhead{R.A. (J2000)} & \colhead{Decl. (J2000)} & \colhead{$z$} & \colhead{$M_{1450,\rm{AB}}$} & \colhead{$A_{\rm V}$} & 
\colhead{Spectrograph}}
\startdata
J$1120$+$0641$  & 11$^\mathrm{h}$20$^\mathrm{m}$01$^\mathrm{s}$.48 & $+$06$^\circ$41\arcmin24\farcs3 & $7.1$ & $-$26.6$\pm$0.1 & 0.1601 &  X-Shooter, FIRE \\ 
J$2348$--$3054$ & 23$^\mathrm{h}$48$^\mathrm{m}$33$^\mathrm{s}$.34 & $-$30$^\circ$54\arcmin10\farcs0 & $6.9$ & $-$25.72$\pm$0.14 & 0.0408 &  X-Shooter  \\
J$0109$--$3047$ & 01$^\mathrm{h}$09$^\mathrm{m}$53$^\mathrm{s}$.13 & $-$30$^\circ$47\arcmin26\farcs3 & $6.7$ & $-$25.52$\pm$0.15 & 0.0669 &  X-Shooter  \\
J$0305$--$3150$ & 03$^\mathrm{h}$05$^\mathrm{m}$16$^\mathrm{s}$.92 & $-$31$^\circ$50\arcmin56\farcs0 & $6.6$ & $-$25.96$\pm$0.06 & 0.0381 &  FIRE \\
\enddata
\end{deluxetable*}    

\subsection{Post-processing}
\label{sec_pprocessing}
In order to homogenize the spectral resolutions and increase the signal to
noise ratio ({\it S/N}) per resolution element, we first smoothed the reduced
spectra by convolving them with a Gaussian kernel with FWHM=100 km s$^{-1}$
($\sigma_{\rm kernel}\sim45$ km\,s$^{-1}$). We then re-sampled the spectra in
velocity space by selecting one representative element every $\Delta v\sim$3
$\sigma_{\rm kernel}$.  Errors on the smoothed spectra were computed from the
errors on the extracted spectra via standard error propagation.  The smoothing
and resampling do not affect our results since we are interested in studying
the continuum and broad emission-line (BEL) properties (typical BEL rest frame
$\sigma_{\rm{line}}>$ 1000 km\,s$^{-1}$).  The spectra were further corrected
for Galactic extinction using the \citet{Cardelli1989} law and $A_{V}$ values
obtained from the extinction map by \citet{Schlegel1998} (assuming $R_V=3.1$,
see Table~\ref{Tab_data}).
 
\begin{figure*}
\centering
\resizebox{0.9\textwidth}{!}{\includegraphics{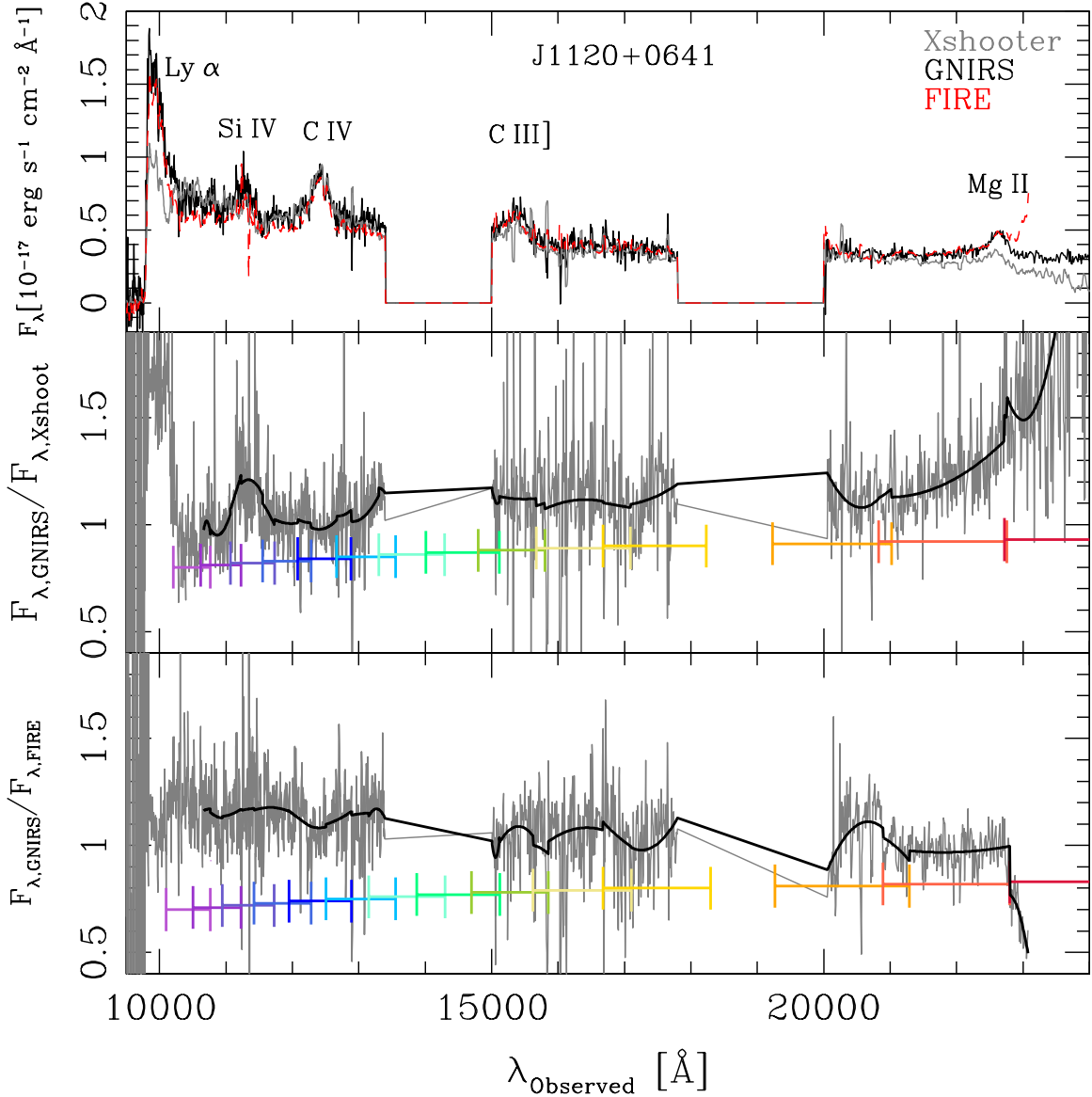}}
\caption{\label{Fig_1} Comparison of the spectroscopic observations of J$1120$+$0641$ -- Top panel: X-Shooter (gray solid line), 
FIRE (red dashed line) and GNIRS data (black solid line). 
Middle \& bottom panels: ratio between the GNIRS and the X-Shooter and FIRE spectra (gray dashed line). 
We have obtained a second-order flux calibration for the 
echellette spectra under the assumption that the GNIRS flux calibration is correct (see text). We have fitted the flux ratio in each echellette order 
(colored segment) with a second order polynomial (black solid line). In case of overlapping orders the final correction is the weighted mean of the individual 
spectral orders. This second-order flux calibration correction can be performed only for J$1120$+$0641$, since this is the only source for which we have non-Echellette 
data.}
\end{figure*}

For J$1120$+$0641$ data were collected with both X-Shooter and FIRE.  The two
reduced spectra present different spectral shapes (see Fig.~\ref{Fig_1}, top
panel), with a clear disagreement between the spectral slopes at both
$\lambda_{\rm{obs}}<11000$ \AA \ and $\lambda_{\rm{obs}}>20000$ \AA. Since
the detected differences in the spectral shape could be driven by a variety of
different causes involving both observations and data reduction (e.g., slit
losses, telluric absorption correction, inter-order flux losses), we further
compared the spectra with the lower resolution spectrum ($R\sim500$) of the
same source obtained with the Gemini near-infrared spectrograph (GNIRS),
published by \citet{Mortlock2011} (see Fig.~\ref{Fig_1}, top panel). GNIRS is
a cross-dispersed spectrograph mounted on the Gemini North Telescope. In
Fig.~\ref{Fig_1} we show the flux ratios between the GNIRS and X-Shooter
spectra (central panel), and between those obtained with GNIRS and FIRE (we
also plot the wavelength ranges of individual orders of X-Shooter and FIRE for
reference).

The X-Shooter spectral shape is overall in good agreement with the GNIRS data,
with the exception of two regions: the \Lya \ region ({\it Y}-band) and the
\Mgii \ region ({\it K}-band). While the {\it Y}-band issues are most probably
related to uncertainties in the absolute flux calibration (the {\it Y}-band
light is measured in part in the VIS arm and in part in the NIR arm), the {\it
  K}-band issues are due to vignetting affecting the redder orders of the NIR
arm.

For the FIRE spectrum, a disagreement with the GNIRS spectral shape is
noticeable, with the difference between the two spectra slowly increasing
towards blue wavelengths and reaching $\sim15$\% in the {\it Y}-band.  While
the differential effect might point towards slit losses, we could not
unequivocally identify the cause of the issues. In addition to the
inconsistencies which vary slowly across the Y to K bands, the FIRE and
X-Shooter data both exhibit order-to-order residuals in the flux ratio,
characteristic of varying {\it S/N} ratio across the echelle blaze in each
order.  Since the low resolution GNIRS spectrum was obtained by adding only 5
different orders (which reduces significantly the fraction of spectrum that
may be affected by inter-order flux losses) and given the overall good
agreement between the GNIRS and X-Shooter spectra, we decided to assume the
GNIRS spectrum to be the best available representation of the source intrinsic
spectrum. We were then able to compute a second-order flux calibration
correction for both the X-Shooter and FIRE spectra. We excluded spectral
regions that were significantly contaminated by telluric absorption (13500 \AA
\ $<\lambda_{\rm{obs}}<$ 15000 \AA, 17800 \AA \ $<\lambda_{\rm{obs}}<$ 20000
\AA) and the \Lya \ region from the computation, given the higher
uncertainties in the flux calibration of the X-Shooter data.  For each
spectral order we fitted the $F_{\lambda,
  \rm{GNIRS}}/F_{\lambda,\rm{X-Shooter,FIRE}}$ flux ratio with a second-order
polynomial.  For overlapping orders, the final correction was computed as the
weighted mean of the individual ones. After applying the second-order flux
calibration corrections, we created a ``final'' J$1120$+$0641$ high {\it S/N}
spectrum by computing the weighted mean of the FIRE and X-Shooter spectra.

\begin{figure*}
\centering
\resizebox{0.9\textwidth}{!}{\includegraphics{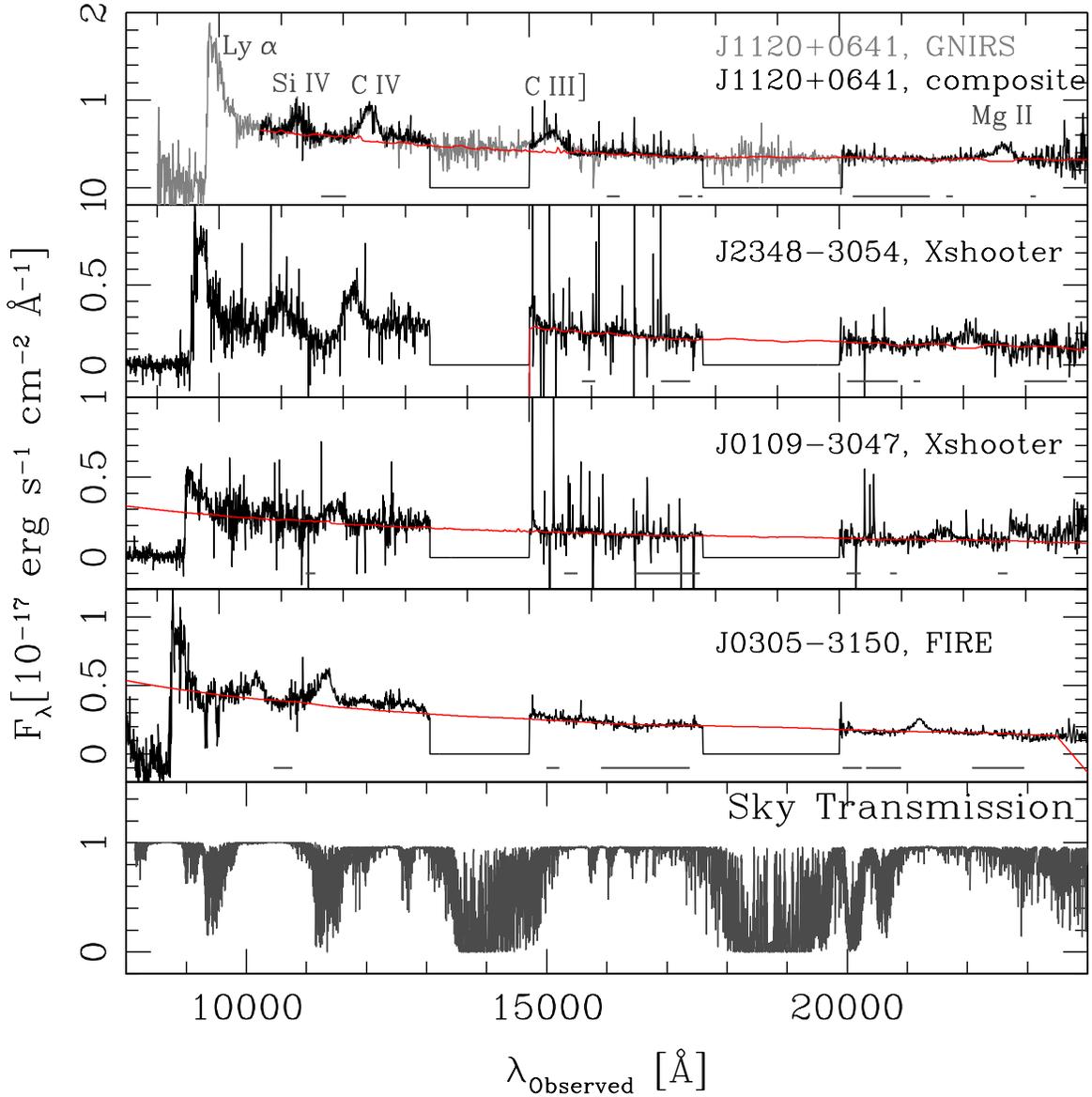}}
\caption{\label{Fig_2} Final optical--to--NIR spectra of all $z>6.5$ quasars. 
Black solid line: observed quasar spectra corrected for Galactic reddening. For J$1120$+$0641$ we plot the composite spectrum  
obtained from the combination of X-Shooter and FIRE data, after secondary flux correction (gray solid line: GNIRS spectrum). 
Red solid line: ``best-fit'' continuum model. For J$2348$--$3054$ (BAL quasar) we limit 
our spectral decomposition to $\lambda_{\rm obs} \geq 15000$ \AA. 
Dark gray segments: continuum 
modeling windows. Bottom panel: Paranal sky transmission spectrum (nominal airmass=1).}
\end{figure*}

It was not possible to compute second-order flux calibration corrections for
J$2348$--$3054$, J$0109$--$3047$ and J$0305$--$3150$ since non-echellette
lower resolution spectra are not available. By multiplying the observed
spectra by the VIKING-VISTA filter throughput curves though, we were able to
compare the broad-band magnitudes obtained from the spectra { (flux calibrated via the spectro-photometric 
standards)} with the catalog ones.  For both the X-Shooter spectra (J$2348$--$3054$ and J$0109$--$3047$)
the derived $Y-J$ color is 0.2 mag redder than the broad--band catalog colors.
For J$0305$--$3150$ instead, the observed $Y-J$ color is in agreement with the
catalog one.  The final spectra are shown in Fig.~\ref{Fig_2}.
 
\hspace{5 pt}

\section{Data Analysis}
We focus on the rest-frame UV spectral region with $1280$ \AA \ $<
\lambda_{\rm rest} < 3000$ \AA.  This wavelength range is characterized by the
presence of strong broad emission lines such as Si\,{\sc iv}+O\,{\sc
  iv}]\,$\lambda \lambda1397,1402$, \Civ\,$\lambda1549$, \CIII\,$\lambda1908$,
and \Mgii\,$\lambda2798$, the non-stellar nuclear continuum, the Balmer
continuum and the \Feii \ and \Feiii \ emission line blends.  The last three
emission features overlap in the spectral range of interest and constitute our
underlying ``pseudo-continuum''.  We excluded from the analysis spectral
regions that are significantly contaminated by telluric absorption (13500 \AA
$<\lambda_{\rm{obs}}<$ 15000 \AA, 17800 \AA $<\lambda_{\rm{obs}}<$ 20000 \AA)
and the \Lya \ region (given the uncertainties that might affect the flux
calibration).  For J$2348$--$3054$, we were forced instead to limit our
analysis to $\lambda_{\rm{obs}}>$ 15000 \AA: this source is a broad absorption
line (BAL) quasar \citep[see also][]{Venemans2013}, hence all the spectral
regions suitable for continuum modeling at 10000 \AA\ $<\lambda_{\rm{obs}}<$
15000 \AA \ are affected by absorption.

\subsection{Spectral model}
\subsubsection{Continuum components}
\label{sec_model_cont}
We modeled the pseudo-continuum with the following components:
\begin{itemize}
\item {\textbf{Non-stellar continuum emission}: the dominant component of a
    quasar spectrum at rest frame UV-optical wavelengths is the non-stellar
    nuclear continuum, modeled as a power-law
\begin{equation}
  F_{\lambda {\rm{,PL}}} = F_{\rm{0,PL}} \left( \frac{\lambda}{1000 \ \rm{\AA}}\right)^\alpha {\rm \  .}
\end{equation}
The determination of the slope coefficient $\alpha$ depends on both the
fitting procedure and the observed spectral range. In the literature, in case
of wide wavelength coverage, the fit of the power-law is usually performed by
selecting narrow fitting windows free of any contribution by other emission
components. { In this work however, in order to increase the width of the
fitting windows and to reduce possible biases due to residual contamination
from other emission features, we model the power-law component simultaneously with
the Balmer continuum and with the \Feii \ emission-line forest. This allows us to adopt 
broader continuum fitting windows (see Fig.\ref{Fig_2})}. 
Lower redshift SDSS quasars ($z<3$) have a typical power-law slope of $\alpha \sim -1.3$, but
the exact value can vary significantly from source to
source. \citet{Decarli2010}, for example, analyzed a sample of 96 quasars at
$z<3$ and obtained a mean value of $\alpha=-1.3$ with a 1-$\sigma$ dispersion
of 1.6. \citet{DeRosa2011} analyzed a sample of 22 high redshift quasars
($4.0<z<6.5$) with NIR spectral coverage ($2000$ \AA \ $\lesssim
\lambda_{\rm{rest}} \lesssim 3000$ \AA) and found a consistent mean value of
$\alpha=-1.5$ with a 1-$\sigma$ dispersion of 1.2.}

\item{\textbf{Balmer continuum emission}: we assume gas clouds with uniform
    temperature $T_{\rm{e}}$, that are partially optically thick. In this
    case, for wavelengths bluer than the Balmer edge ($\lambda_{\rm{BE}}=3646$
    \AA, rest frame), the Balmer spectrum can be parameterized as
    \citep{Grandi1982,Wills1985}
\begin{equation}
 F_{\lambda}=F_{\rm{BE}} \ B_{\lambda}(T_{\rm{e}}) \ (1-\rm{e}^{-\tau_{\rm{BE}}({\lambda}/{\lambda_{\rm{BE}}})^3}), \ \lambda<\lambda_{\rm{BE}}  {\rm \  ,}
\end{equation}
where $B_{\lambda}(T_{\rm{e}})$ is the Planck function at the electron
temperature $T_{\rm{e}}$, $\tau_{\rm{BE}}$ is the optical depth at the Balmer
edge, and $F_{\rm{BE}}$ is the normalized flux density at the Balmer edge.  We
have used the library of Balmer continuum emission templates created by
\citet{Dietrich2003}. The library includes 16 Balmer continuum templates
probing different values of the (a) electron temperature
($T_{\rm{e}}=10000,12500,15000,20000$ K) and (b) optical depth at the Balmer
edge ($\tau_{\rm{BE}}=0.1,0.5,1,2$).
We consider each template as a fixed point in our multidimensional grid and evaluate the template likelihood as a function of the 
template normalization, $F_{\rm{0,BC}}$, i.e.,    
\begin{equation}
 F_{\lambda{\rm{,BC}}} = F_{\rm{0,BC}} \ F_{\rm{template,BC}}(T_{\rm{e}},\tau_{\rm{BE}})  {\rm \  .}
\end{equation}
}

\item{{ Fe\,{\footnotesize\bf II} and Fe\,{\footnotesize\bf III} emission
      lines}: The \Feii \ and \Feiii \ ions emit a forest of lines, many of
    which are blended.  We model the \Feii \ + \Feiii \ emission in our quasar
    spectra using a scaled and broadened version of the empirical emission
    template spectrum of \citet{Vestergaard2001}. This template is based on
    the high resolution spectrum of the narrow line Seyfert 1 galaxy
    PG0050+124 ($z=0.061$), observed with the {\it Hubble Space
      Telescope}. Vestergaard \& Wilkes isolated the emission-line template by
    fitting and subtracting the power-law continuum and the absorption
    emission features from all the elements but Fe. Since we are mainly
    interested in the \Feii/\Mgii \ flux ratio, we model the continuum on
    windows characterized by \Feii \ complexes only. The intrinsic broadness
    of the \Feii \ component, $\sigma_{\rm{QSO}}$, is unknown and varies from
    quasar to quasar, therefore we have created a library of \Feii + \Feiii \
    templates by convolving the original template in velocity space with a
    Gaussian kernel with
\begin{equation}
\sigma=\sqrt{\rm{FWHM}^2_{\rm{QSO}}-\rm{FWHM}^2_{\rm{Templ}}} \ / \ 2 {\rm ln} (2\sqrt{2})  {\rm \  ,}
\end{equation}
where $\rm{FWHM}_{\rm{Templ}}$ is the characteristic FWHM of iron features in
the spectrum of PG0050+124, and the final FWHM of the iron features in the
quasar spectrum varies between $\rm{FWHM}_{\rm{QSO}}=(1000$ - $16000)$ km
s$^{-1}$, in steps of $500$ km s$^{-1}$.  We consider each template as a fixed
point in our multidimensional grid and evaluate the template likelihood as a
function of the template normalization, $F_{\rm{0,Fe}}$,
\begin{equation}
 F_{\lambda {\rm{,Fe}}} = F_{\rm{0,Fe}} \  {\rm Fe}_{\rm{\ Templ}}(\sigma)  {\rm \  .}
\end{equation}
} 
\end{itemize}

\subsubsection{Emission line components}
\label{sec_model_lines}
In low {\it S/N} regimes { (average continuum {\it S/N} per pixel $\lsim$10)}, details of the emission-line profiles 
are not discernible, but we find that the emission lines can be well modeled by simple
Gaussian functions
\begin{equation}
F_{\lambda,\rm{G}} =  F_{\rm{0,G}} \ {\rm exp} \left({\frac{(\lambda - \lambda_0)^2}{2 \sigma^2}}\right)  {\rm \  ,} 
\end{equation}
where the line amplitude $F_{\rm{0,G}}$, the line dispersion $\sigma$ and the
peak wavelength $\lambda_0$ are the model parameters.  At higher {\it S/N} 
{ (average continuum {\it S/N}  per pixel $\gsim$10)} the
level of detectable details in the line shape improves significantly: emission
lines often present asymmetric profiles characterized by prominent red and/or
blue wings, and various degrees of kurtosis. Therefore, in high {\it S/N}
regimes, the simple Gaussian description often becomes unsatisfactory.  In
cases where the emission line is similar to a Gaussian but asymmetric,
\citet{Vandermarel1993} show that the line can be well described by a
particular subset of Gauss-Hermite polynomials of fourth degree \citep[see
also][]{Riffel2010}
\begin{equation}
{F}_{\lambda,{\rm{GH}}} =  {F}_{{0,{\rm{GH}}}} \frac{e^{-w^2/2}}{\sigma \sqrt{2\pi}} [1+h_3 H_3(w)+h_4 H_4(w)]  {\rm \  ,}  
\end{equation}
where $F_{\rm{0,GH}}$ is the amplitude of the Gauss-Hermite series,
$w=(\lambda-\lambda_0)/\sigma$, $h_j$ are the Gauss-Hermite moments and the
Hermite polynomials $H_j(w)$ are respectively
\begin{equation}
H_3(w)=\frac{1}{\sqrt{6}}(2\sqrt{2}w^3-3\sqrt{2}w) 
\end{equation}
and
\begin{equation}
H_4(w)=\frac{1}{\sqrt{24}}(4w^4-12w^2+3) {\rm \  .} 
\end{equation}
The $h_3$ Gauss-Hermite moment represents the degree of asymmetry with respect
to a Gaussian profile, while the $h_4$ moment assesses the degree of kurtosis
or ``peakiness'' of the line ($h_4>0$: profile more boxy than Gaussian
profile; $h_4<0$: profile more peaky than Gaussian profile). In case
$h_3=h_4=0$ $F_{\lambda,\rm{GH}}$ is equal to a simple Gaussian profile.  
{ We fitted each of the detected bright emission lines (\SiIV, \CIII, \Civ, and
\Mgii) with a Gaussian profile for the sources characterized by a continuum {\it S/N}$\lsim$10 
(J$0109$-$3047$ and J$2348$--$3054$) and with a Gauss-Hermite polynomial for the 
remaining sources (J$1120$+$0641$ and J$0305$--$3150$)}. Note that, given the
spectral quality and the final resolution of our spectra:
\begin{itemize}
\item { the individual transitions of O\,{\sc iv}] and the \SiIV, \Civ, and \Mgii \ 
doublets are blended. Therefore we did not attempt to isolate their individual components 
\citep[see also][]{Jiang2007};}
\item { we ignored Al\,{\sc iii}\,$\lambda1857$ and Si\,{\sc iii}]\,$\lambda1892$ emission lines that are blended with the strong
\CIII \ emission since the latter is typically significantly brighter \citep[e.g.][]{Vanden2001}; 
and}
\item { we did not attempt to model any narrow emission line component, since it is not possible to 
reliable subtracting them in presence of strong broad emission lines without unblended templates for the 
velocity widths \citep[that do not exist, e.g.][]{Denney2014}}
\end{itemize}

\subsection{Spectral modeling}
\label{sec_procedure}
We perform the spectral modeling in the source rest-frame. For high redshift
quasars, it is possible to obtain a direct estimate of the systemic redshift
through the detection of cold molecular gas phase in the host-galaxy
\citep[e.g.,][]{Walter2003,Wang2010, Venemans2012}.  \citet{Venemans2012}
observed the the \CIIforb\,$\lambda158\,\micron$ emission line for
J$1120$+$0641$ and measured a systemic redshift
$z_{\rm{[CII]}}=7.0842\pm0.0004$, which is in agreement with the redshift
estimate obtained by \citet[][$z=7.085\pm0.003$]{Mortlock2011} from the
cross-correlation of the observed \Mgii \ emission line with the quasar
composite spectrum of \citet{Hewett2010}. Therefore, for J$1120$+$0641$, we
fixed the systemic redshift to the $z_{\rm{[CII]}}$ estimate.  For
J$0305$--$3150$, J$0109$--$3047$ and J$2348$--$3054$ there are no available
observations of the cold molecular gas phase.  Therefore, for these objects,
we assume that the \Mgii \ emission line (which is a low ionization line) as a
proxy of the quasar systemic redshift \citep[see
also][]{Jiang2007,Kurk2007,DeRosa2011}. Low-ionization lines are preferred to
high-ionization lines (e.g., \Civ) because the latter can present
high-velocity offsets with respect to the source systemic redshift
\citep[$\Delta {\rm{v}} \gtrsim1000$ km s$^{-1}$, corresponding to 
$\Delta z\sim0.02$ at $z\sim6$, see, e.g.,][ and references therein]{Richards2002,Richards2011,Shang2007}.
Hence, for the VIKING quasars, we assumed as initial guess for the systemic
redshift the one obtained from the onset of the \Lya \ Gunn-Peterson
absorption \citep{Venemans2013}. The initial guess was then substituted by
$z_{\rm{MgII}}$, and both continuum and line modeling were iteratively
performed till the convergence on $z_{\rm{MgII}}$ was reached.

The modeling of the continuum is performed following a grid-based
approach. Given the continuum model,
\begin{equation}
{F}_{\lambda,{\rm{Model}}} =  F_{\lambda {\rm{,PL}}} + F_{\lambda{\rm{,BC}}} +  F_{\lambda {\rm{,Fe}}} {\rm  ,}
\end{equation}
we computed the likelihood of the data { in the spectral regions that show only continuum and \Feii \ emission}
 ($d_{\rm{cont}}$) given the model parameters 
($\Theta$) as $P(d_{\rm{cont}} | \Theta) = N e^{-\chi^2/2}$, where $N$ ensures that the
likelihood is normalized.  
In the likelihood definition, we have
\begin{equation}
\chi^2 \ = \ \sum_{i} \frac{({F}_{\lambda ,{\rm obs},i}-{F}_{\lambda,{\rm Model},i})^2}{\sigma_i^2} {\rm  ,}
\end{equation}
where ${F}_{\lambda\,{\rm obs},i}$ is the observed flux, $\sigma$ is the
uncertainty on the observed flux and the index $i$ runs over the pixels that
present only continuum and \Feii \ emission.  This definition of the
likelihood assumes that the uncertainty on the flux in individual pixels is
Gaussian.

Using a $3 \sigma$-clipping algorithm, we automatically excluded noise peaks
and residuals from telluric absorption correction from our fitting windows.
There are six model parameters $\Theta$: power-law normalization and slope,
Balmer continuum normalization and template identifier, iron template
normalization and identifier. All the parameters, with the exception of the
Balmer continuum and iron template identifiers, where discretized over a
regular grid. The parameter ranges
were chosen to ensure that each of the marginal probability
distributions goes to zero at the boundaries of the probed interval.
We assumed priors that are uniform over the specified domain for $\Theta$, 
and zero elsewhere. This implies that, within the domain, the posterior
probability distribution function (pdf) is proportional to the likelihood,
$P(\Theta|d_{\rm{cont}}) \propto P(d_{\rm{cont}}|\Theta)$. 
The resolution of the individual grid was chosen and adapted in
order to properly sample the peak of the posterior distribution.  
Hereafter we indicate the continuum model with the
maximum posterior pdf (or maximum likelihood) as the ``best-fit'' continuum
model.

{ To derive the emission lines properties we assumed either a Gaussian or a Gauss-Hermite line-profile 
(see sec.~\ref{sec_res_lines}). 
We indicate the set of unknown parameters of a line with $\Lambda$. 
After subtracting the ``best-fit'' continuum model, we derived $\Lambda$
using a $\chi^2$ minimization routine. We then measured the line
FWHM and dispersion \citep[following][]{Peterson2004}, the wavelength
corresponding to the line peak ($\lambda_{\rm{Peak}}$), the line flux, 
and the EW directly from the line model. We adopted the corresponding values as
``best-fit'' estimates of the line properties.}

{ The vast majority of the studies in the literature do not take into account the errors on the emission 
line properties that result from the uncertainties of the continuum modeling. This is particularly important when the {\it S/N} of the 
spectra is poor and degeneracies among the model components are present. 
For these reasons, we estimated the errors on the line properties in two distinct steps.
First, we estimated the uncertainty on $\Lambda$, given the posterior 
distribution of the continuum parameters, $P(\Theta|d_{\rm{cont}})$.
This accounts for the impact of potential degeneracies in the continuum parameters.
Second, we estimated the uncertainty due to the {\it  S/N} of the line, for a fixed continuum model 
(fixed $\Theta$).

In the first step, we sampled $\Theta$ from $P(\Theta|d_{\rm{cont}})$ with a Monte Carlo rejection
method. For each sample we subtracted the corresponding continuum model. We then 
restricted the analysis to the continuum subtracted data within a line-fitting spectral region 
($d_{\rm{line}}$) 
 to obtain the estimated line properties from the $\chi^2$ minimization routine 
($\Lambda_{\rm best-fit}(\Theta)$). 
This is equivalent to obtaining the pdf of the line properties, $P(\Lambda|d_{\rm{line}})$,
marginalized over the distribution of the continuum models
\begin{equation}
\label{PP}
 P_\delta (\Lambda|d_{\rm{line}}) = \int \delta(\Lambda - \Lambda_{\rm best-fit}(\Theta)) \ P(\Theta | d_{\rm{cont}}) {\rm d}\Theta {\rm  ,}
\end{equation}
under the assumption that $P(\Lambda|\Theta, d_{\rm{line}})$ is a Dirac's $\delta$ function
centered on the line best-fit properties, i.e., $P(\Lambda | \Theta,d_{\rm{line}}) = 
\delta(\Lambda - \Lambda_{\rm best-fit}(\Theta))$. 
One can think of $P_\delta (\Lambda|d_{\rm{line}})$ as a weighted mean of the best-fit line properties at fixed
$\Theta$, with weights proportional to $P(\Theta|d_{\rm{cont}})$.

However, for a given continuum model (for given $\Theta$), 
the line property estimates are significantly affected by sources of error that depend
not only on the overall {\it S/N} of the data, but also on the assumed line profile, on
intrinsic absorption lines and on telluric contamination. This means that
$P(\Lambda|\Theta,d_{\rm{line}})$ is not a $\delta$ function.

The second step of our error estimate consists in evaluating the spread of
$P(\Lambda | \Theta,d_{\rm{line}})$.}  To do so, we followed \citet{Assef2011} and used a
Monte Carlo approach.  Starting from the line best-fit model and from the
measured flux uncertainty, we generated 5000 resampled spectra where the flux
in each pixel was randomly drawn from a Gaussian distribution with mean value
equal to the best-fit model flux and dispersion equal to the flux uncertainty.
By re-measuring the line properties for each resampled spectrum we obtained
$P(\Lambda | \Theta,d_{\rm{line}})$.  This distribution is well approximated by a Gaussian
of which we evaluated the dispersion $\sigma_{P(\Lambda | \Theta,d_{\rm{line}})}$.  After
checking that the dispersion does not depend strongly on the choice of the
subtracted continuum ($\sigma_{P(\Lambda | \Theta, d_{\rm{line}})} \simeq \sigma_{P(\Lambda | d_{\rm{line}})} \simeq constant \equiv
\sigma_{\Lambda}$) we obtained our line parameter pdf ($P(\Lambda|d_{\rm{line}})$) by convolving
$P_\delta (\Lambda|d_{\rm{line}})$ with a Gaussian distribution having dispersion equal to
$\sigma_{\Lambda}$.  This is equivalent to substituting the
$\delta(\Lambda - \Lambda_{\rm best-fit}(\Theta))$ in Eq.~\ref{PP} with $N(\mu=\Lambda_{\rm best-fit}(\Theta),\sigma_{\Lambda}^2)$.

From this final pdf we estimated the confidence level of each line property as
the interval in which the cumulative pdf goes from 16\% to 84\% (the central 68\% 
credible interval, equivalent to
a 1$\sigma$ confidence level for a Gaussian pdf).

\section{Results}

\subsection{Quasar continuum modeling}
\begin{figure*}
\centering
\resizebox{0.7\textwidth}{!}{\includegraphics{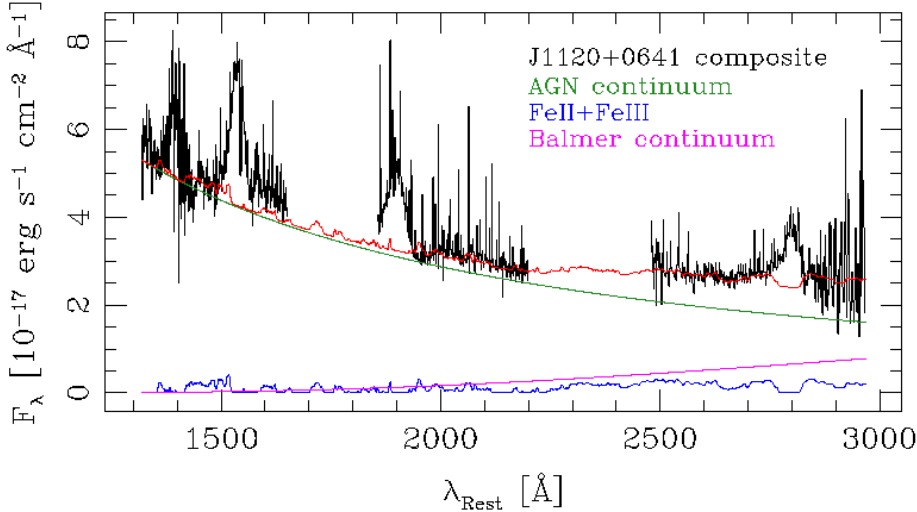}}
\caption{\label{Fig_3} Continuum spectral decomposition for J$1120$+$0641$: AGN continuum (green), 
Balmer continuum (magenta), \Feii + \Feiii \ line forest (blue), pseudo-continuum ``best-fit'' model (red) corresponding to the sum of the AGN continuum, the Balmer continuum, and the  \Feii + \Feiii \ line forest.}
\end{figure*}

The ``best-fit'' continuum models are shown in Fig.~\ref{Fig_2} (red solid
lines). In Fig.~\ref{Fig_3} we show the ``best-fit'' continuum components
obtained for J$1120$+$0641$ as an example of spectral decomposition.  Given
the individual shapes of the power-law continuum and of the Balmer continuum
in the spectral range of interest, the results for the contributions of these
two components are expected to be strongly correlated. { Therefore it is important 
to simultaneously model all the pseudo continuum components in order to properly account 
for degenaracies.}
In Table~\ref{Tab_measure} we list the best-fit estimates of the power-law slope
parameter together with the values corresponding to the 1$\sigma$ confidence
level. The slope coefficients are in agreement within 1$\sigma$ with both the
local slope estimated by \citet{DeRosa2011} and the global value obtained by
\citet{Decarli2010}.

We estimated the uncertainties on the individual parameters by using the
corresponding marginal pdfs.  While for J$1120$+$0641$ the marginal pdfs for
the continuum parameters are regular and show clear individual peaks, { for the
VIKING quasars we observe very low (if any) contribution of the Balmer continuum. 
This implies that the Balmer continuum model is not well constrained. 
Moreover, for J$0109$--$3047$ we cannot
exclude a zero contribution of the iron component to the continuum model (see
Section~\ref{sec_Fe} for a more detailed discussion)}.

\begin{deluxetable*}{lcccc}[t]
\tabletypesize{\footnotesize}
\tablecaption{\label{Tab_measure} Modeled spectral properties.}
\tablecolumns{5}
\tablehead{
\colhead{} & \colhead{J$1120$+$0641$} & \colhead{J$2348$--$3054$} & \colhead{J$0109$--$3047$} & \colhead{J$0305$--$3150$}}
\startdata
$\alpha$ & $-1.47^{+0.02}_{-0.01}$ & $-2.56^{+0.51}_{-0.63}$ & $-1.33^{+0.22}_{-0.07}$ & $-1.37^{+0.01}_{-0.01}$ \\
$z_{\rm MgII}$ & $7.097^{+0.002}_{-0.001}$ & $6.889^{+0.007}_{-0.006}$ & $6.747^{+0.007}_{-0.005}$ & $6.605^{+0.002}_{-0.001}$ \\
$z_{\rm SiIV}$ & $7.068^{+0.002}_{-0.003}$ & -- & $6.732^{+0.013}_{-0.010}$ & $6.595^{+0.002}_{-0.002}$ \\
$z_{\rm CIV}$ & $7.024^{+0.001}_{-0.001}$ & -- & $6.675^{+0.006}_{-0.002}$ & $6.573^{+0.001}_{-0.004}$ \\
$z_{\rm CIII]}$ & $7.0387^{+0.0007}_{-0.0007}$ & -- & -- & -- \\
$\lambda L_{\lambda}(1350 \ {\rm \AA})$ $[10^{46} \ {\rm erg} \ {\rm s}^{-1}]$ & $4.20 \pm 0.02$ & -- & $1.39\pm0.06$ & $2.22 \pm 0.02$ \\
$\sigma_{\rm CIV} \ [{\rm \AA}]$ & $14.72^{+0.15}_{-0.17}$ & -- & $16.5^{+1.1}_{-2.5}$ & $18.63^{+0.23}_{-0.62}$ \\
$\lambda L_{\lambda}(3000 \ {\rm \AA})$ $[10^{46} \ {\rm erg} \ {\rm s}^{-1}]$ & $2.90 \pm 0.03$ & $0.94 \pm 0.33$ & $1.07 \pm 0.14$ & $1.66\pm0.02$ \\
${\rm FWHM}_{\rm MgII} \ [{\rm km \ s}^{-1}]$ & $4411^{+210}_{-150}$ & $5446^{+580}_{-360}$ & $4389^{+400}_{-360}$ & $3189^{+110}_{-60}$ \\
$F_{\rm SiIV} \ [10^{-17}  \ {\rm erg} \ {\rm s}^{-1} \ {\rm cm}^{-2}]$ & $41.30^{+0.15}_{-0.17}$ & -- & $13.1^{+6.2}_{-3.8}$ & $37.4^{+0.9}_{-1.4}$ \\
$F_{\rm CIV} \ [10^{-17}  \ {\rm erg} \ {\rm s}^{-1} \ {\rm cm}^{-2}]$ & $117.0^{+1.2}_{-0.8}$ & -- & $33.9^{+6.4}_{-3.2}$ & $72.4^{+1.1}_{-1.4}$ \\
$F_{\rm CIII} \ [10^{-17}  \ {\rm erg} \ {\rm s}^{-1} \ {\rm cm}^{-2}]$ & $85.6^{+0.7}_{-0.7}$ & -- & -- & -- \\
\enddata
\end{deluxetable*}

{ We point out that, since for the VIKING quasars we were not able to properly cross-check the
goodness of the flux calibration and to compute appropriate flux calibration
corrections (see Section~\ref{sec_pprocessing}), continuum decomposition results hold valid as long as 
the global spectral shape is preserved.}

\subsection{Emission line modeling and redshift estimates}
\label{sec_res_lines}
\begin{figure*}
\centering
\resizebox{0.5\textwidth}{!}{\includegraphics{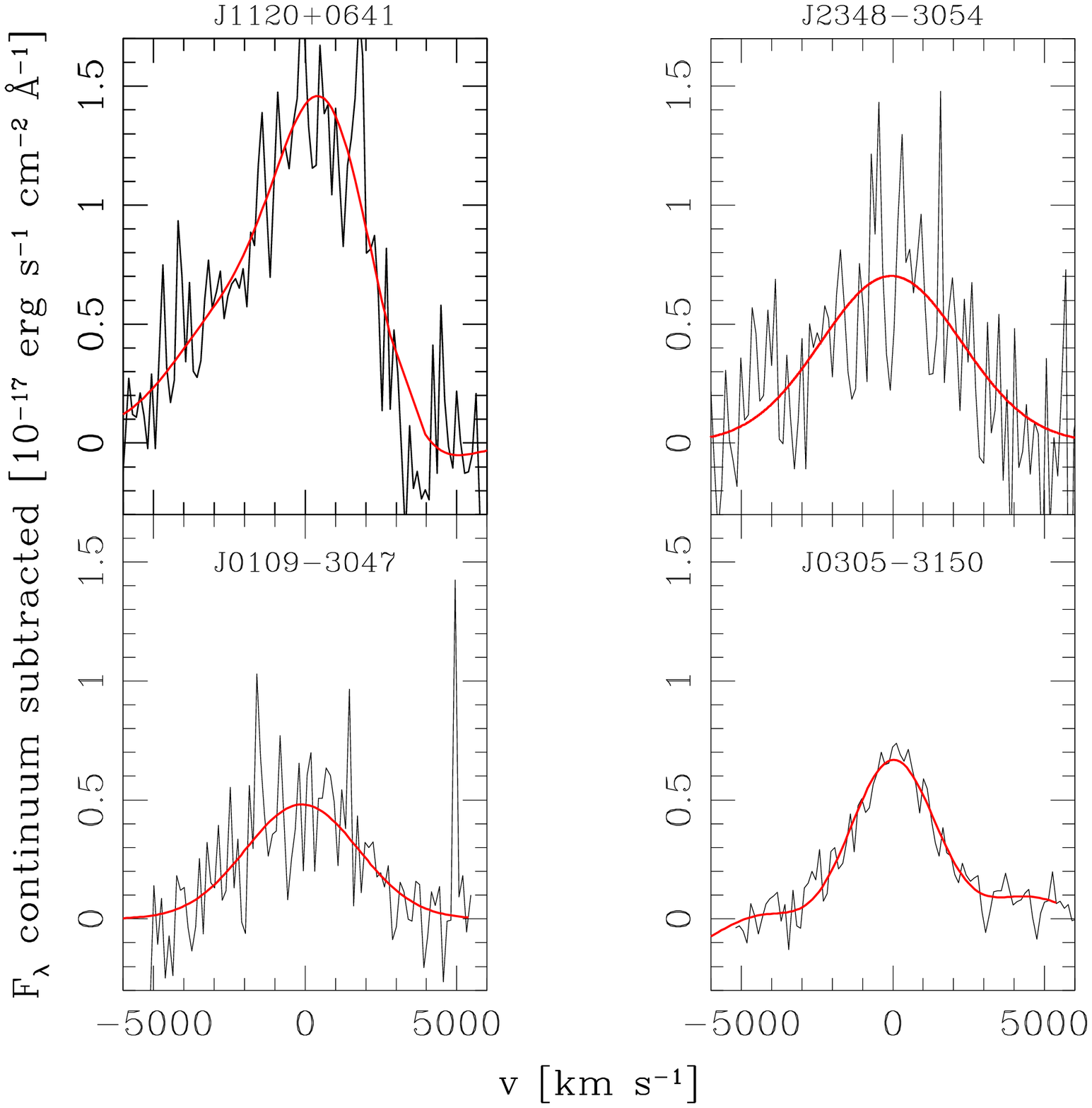}}
\caption{\label{Fig_4} Black solid line: rest frame flux density in the \Mgii
  \ fitting window after subtraction of the ``best-fit'' continuum model. The
  spectra have been redshifted to the rest frame system of reference by using
  the source nominal redshift ($z_{\rm{J1120}}=$7.084, $z_{\rm{J2348}}=$6.89,
  $z_{\rm{J0109}}=$6.75, $z_{\rm{J0305}}=$6.606). { Red solid line: ``best-fit'' 
  emission line model}.}
\end{figure*}

\begin{figure*}
\centering
\resizebox{0.9\textwidth}{!}{\includegraphics{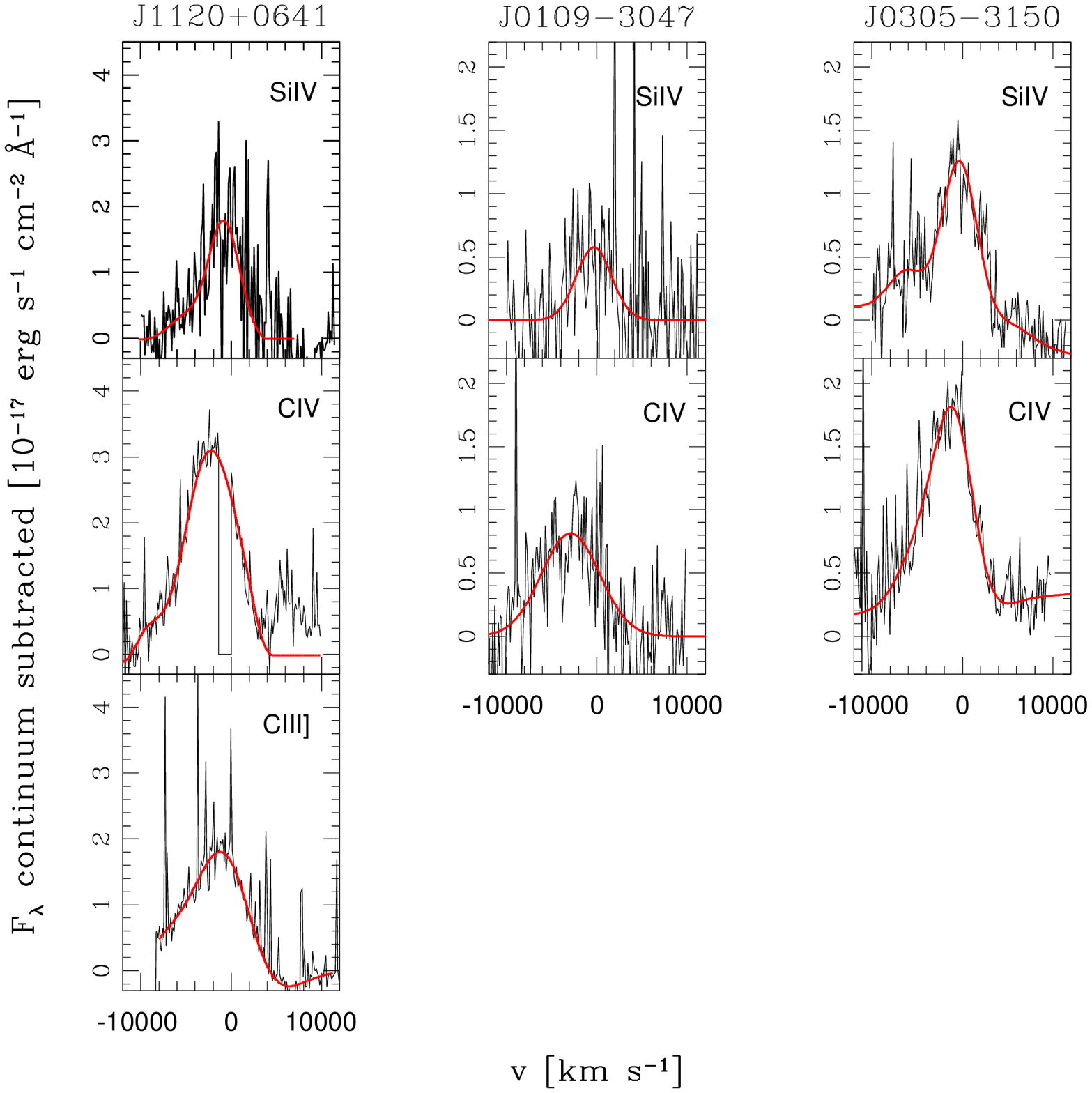}}
\caption{\label{Fig_5} \SiIV, \CIII, and \Civ \ emission lines. Given the
  systemic redshift of J$0109$--$3047$ and J$0305$--$3150$, it was not
  possible to model their \CIII \ emission line due to telluric
  contamination. The spectra have been redshifted to the rest frame system of
  reference by using the source nominal redshift ($z_{\rm{J1120}}=$7.084,
  $z_{\rm{J0109}}=$6.75, $z_{\rm{J0305}}=$6.606).  Black solid line: rest
  frame flux density after subtraction of the ``best-fit'' continuum model. 
  { Red solid line: ``best-fit'' emission line model.}
  Note that, for both J$1120$+$0641$ and J$0305$--$3150$, \SiIV, \CIII, and
  \Civ \ are blue-shifted with respect to their nominal central wavelengths.}
\end{figure*}

\begin{figure*}
\centering
\resizebox{0.5\textwidth}{!}{\includegraphics{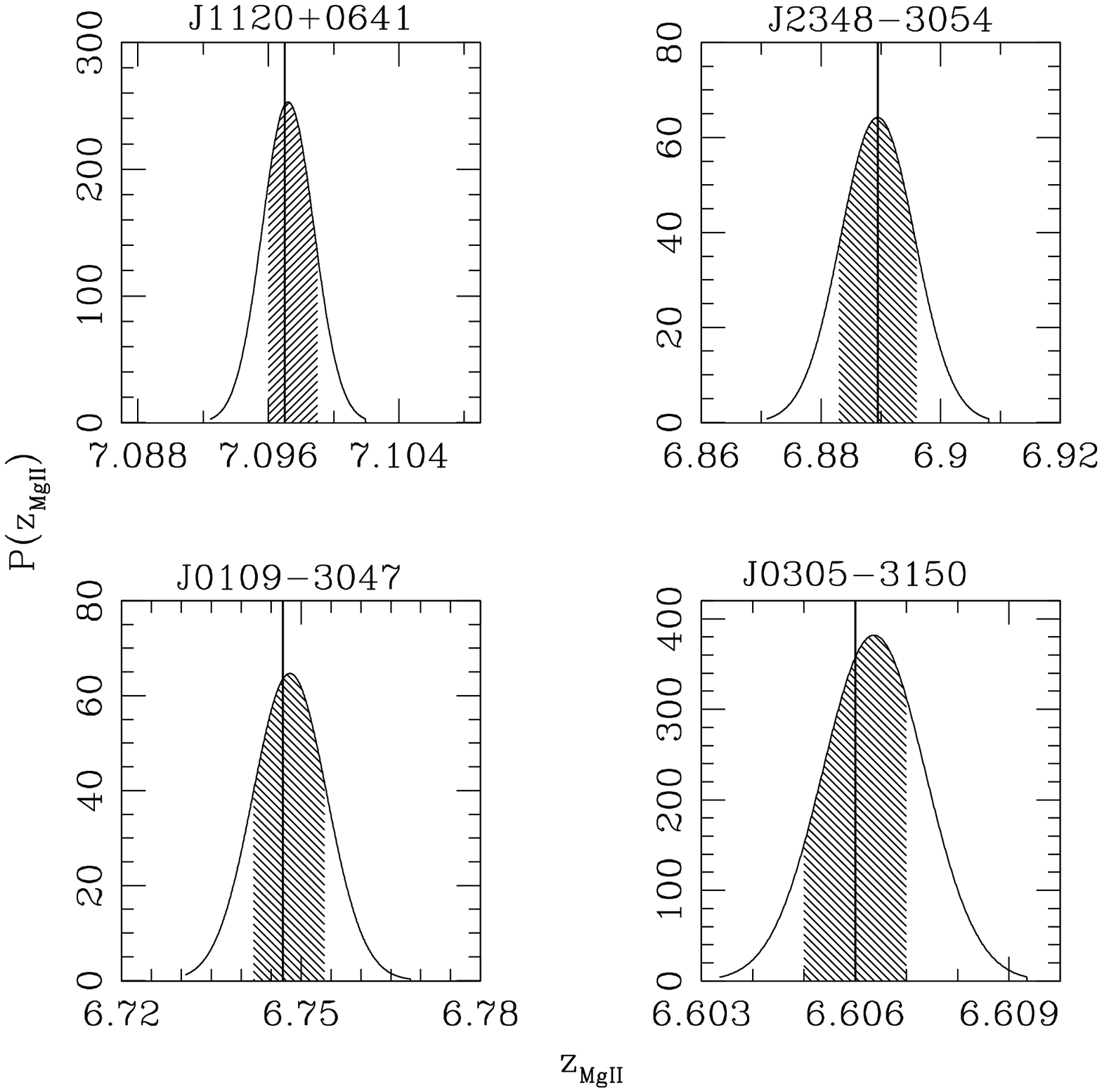}}
\caption{\label{Fig_6} { \Mgii \ redshift: marginal
  probability distributions.  The vertical lines indicate the ``best-fit''
  estimate, while the shaded areas correspond to the 1$\sigma$ confidence
  level (see Section~\ref{sec_procedure} for details).}}
\end{figure*}

We modeled all the strong UV emission lines detected in the spectrum of
J$1120$+$0641$: \SiIV, \Civ, \CIII, and \Mgii.  Given the systemic redshift
of J$0109$--$3047$ and J$0305$--$3150$, it was not possible to model their
\CIII \ emission line due to telluric contamination. Finally, for
J$2348$--$3054$, we fully modeled only the \Mgii \ emission line since both
\SiIV \ and \Civ \ emission lines are severely affected by absorption (this
source is a BAL quasar).  Emission-line models obtained after the subtraction
of the ``best-fit'' continuum model are shown in Fig.~\ref{Fig_4} and
Fig.~\ref{Fig_5}. For each line, we estimated the redshift as
\begin{equation}
\label{eq_z}
 z_{\rm line}+1=\frac{\lambda_{\rm 0,line}}{\mu_{\rm line}} {\rm ,}
\end{equation}
where $\lambda_{\rm 0,line}$ is the wavelength of the line peak measured in
the observed frame and $\mu_{\rm line}$ is the reference laboratory wavelength
for the line transition. { In Table~\ref{Tab_measure} we list the obtained redshifts
while in Fig.~\ref{Fig_6} we show, as an example, the marginal probability distributions 
for the \Mgii \ line}.


{ For J$1120$+$0641$, the $z_{\rm MgII}$ estimate and the systemic redshift 
obtained from molecular gas
\citep{Venemans2012} disagree by more than 3$\sigma$}, suggesting a red-shift
of the \Mgii \ low ionization line. On the other hand, the \SiIV, \Civ, and
\CIII \ emission lines for this source present significant blue-shifts with
respect to the source systemic redshift.  Analogously, for J$0305$--$3150$,
both \SiIV \ and \Civ \ emission lines are significantly blue-shifted with
respect to the \Mgii \ emission line, that we adopt as a proxy for the
systemic redshift (see Section~\ref{sec_procedure}).  However, we need to
stress that our estimates of the blue-shifts for the \SiIV \ and \CIII \
emission lines might be partially affected by the fact that, given the {\it
  S/N} of our spectra, we model these systems as individual transitions,
ignoring the blended components (see Section~\ref{sec_model_lines}).  It is
also worth noticing that the error estimates we cite do not include systematic
uncertainties, such as the choice of the line fitting intervals. { For example,
for the \Civ \ emission line of J$1120$+$0641$, small differences in the
choice of the wavelengths excluded from the modeling due to the strong absorption feature
 affecting the line peak (see Fig.~\ref{Fig_5}, left column, central panel)}, can cause the redshift to
vary up to $\Delta z \sim 0.01$ ($\Delta v\sim 350$ km s$^{-1}$).

For both J$1120$+$0641$ and J$0305$--$3150$, the \Civ \ emission line presents
the largest blue-shifts with $\Delta v_{\rm J1120+0641} = 2260\pm40$ km
s$^{-1}$ and $\Delta v_{\rm J0305-3150} = 1300\pm150$ km s$^{-1}$. For the
\Civ \ emission line in J$1120$+$0641$, \citet{Mortlock2011} measured a
blue-shift of $\Delta v=2800\pm250$ km s$^{-1}$. While the two estimates seem
to disagree by more than 2$\sigma$, they should not be directly compared since
the emission-line shifts have been estimated in two different ways. We
estimated the line redshift starting from the wavelength of the line peak
$\lambda_{\rm 0,line}$ (see Eq.~\ref{eq_z}), while \Citet{Mortlock2011}
estimated it starting from a flux weighted central wavelength.
       
\subsection{Black Hole Masses}
\label{sec_MBH}
The width of broad emission lines detected in the spectra of AGN is thought to
originate primarily from Doppler broadening related to the motion of the
emitting gas around the central BH. Under the assumption that the emitting gas
is virialized, \Mbh \ can be obtained by combining an estimate of the gas
distance from the BH (broad line region radius, $R_{\rm BLR}$) with the
velocity of the clouds emitting at $R_{\rm BLR}$ ($v_{\rm BLR}$),
\begin{equation}
 M_{\rm{BH}}= f  \ {G}^{-1} \ R_{\rm BLR} \ v^2_{\rm BLR} {\rm ,}
\end{equation}
where the factor $f$ is a scale factor that depends on inclination, geometry
and kinematics of the BLR \citep[e.g.][]{Peterson1999,Decarli2008, Grier2013}
and $\rm G$ is the gravitational constant. The product $G^{-1} \ R_{\rm BLR} \
v^2_{\rm BLR}$ includes all the observables, and is often referred to as
"virial product".  While the gas velocity can be obtained from the width of the
broad emission lines, the only way to directly estimate $R_{\rm BLR}$ in
non-nearby AGN is through reverberation mapping \citep[RM;
e.g.,][]{Peterson2004}.  RM studies of \Hb \ emission in local AGNs have lead
to the discovery of a tight correlation between the AGN continuum luminosity
and the distance of the emitting gas \citep[$R_{\rm BLR}-L$
relation;][]{Kaspi2005,Bentz2009,Bentz2013,Zu2011}.  Thanks to this relation,
it is possible to estimate the BLR size, and consequently \Mbh, from single
epoch spectra.  For high redshift sources, the \Hb \ emission line is
redshifted out of the visible window, and \Mbh \ estimates are then based on
the \Civ \ and \Mgii \ emission line. For mass estimates based on \Civ \ we
used the empirical \Mbh \ calibration obtained by \citet{Vestergaard2006}
\begin{equation}
  M_{\rm BH} (\mbox{C\,\sc{iv}})=10^{6.73} \left( \frac{\sigma_{\rm line}}{10^3 \ {\rm km} \ {\rm s}^{-1}} \right)^2 \left( \frac{\lambda L_\lambda(1350 {\rm \ \AA})}{10^{44} \ {\rm erg \ s}^{-1}} \right)^{0.53} M_\odot {\rm ,}
\end{equation}
where $\sigma_{\rm line}$ is the \Civ \ line dispersion and $\lambda
L_\lambda(1350 {\rm \ \AA})$ is the monochromatic AGN continuum luminosity
estimated at $\lambda_{\rm rest}=1350$ \AA (see Table~\ref{Tab_measure}). 
For mass estimates based on \Mgii
\ instead, we used the empirical calibration obtained by
\citet{Vestergaard2009}
\begin{equation}
  M_{\rm BH} (\mbox{Mg\,\sc{ii}})=10^{6.86} \left( \frac{\rm FWHM_{\rm line}}{10^3 \ {\rm km} \ {\rm s}^{-1}} \right)^2 \left( \frac{\lambda L_\lambda(3000 {\rm \ \AA})}{10^{44} \ {\rm erg \ s}^{-1}} \right)^{0.5} M_\odot {\rm ,}
\end{equation}
where ${\rm FWHM}_{\rm line}$ is the \Mgii \ line FWHM and $\lambda
L_\lambda(3000 {\rm \ \AA})$ is the monochromatic AGN continuum luminosity
estimated at $\lambda_{\rm rest}=3000$ \AA (see Table~\ref{Tab_measure}). Both relations were calibrated to
the \citet{Peterson2004} RM results, and are consistent within $\sim$0.1 dex
\citep{Vestergaard2009}.  The 1$\sigma$ scatter in the absolute zero points
are equal to 0.32 dex for the \Civ \ relation and to 0.55 dex for the \Mgii \
relation.  The intrinsic scatter of the estimators dominates the \Mbh \
measurement uncertainties.  { For the \Civ \ emission line we opted to estimate the gas
velocity from the line dispersion instead of the line FWHM, following 
\citet{Denney2013}.}
\begin{deluxetable*}{lllll}
  \tablecolumns{5} 
\tablewidth{0pc} 
\tablecaption{\label{Tab_derive} Estimated \Mbh, quasar Eddington ratios, emission line ratios and \Civ \ EW.}
  \tablehead{ \colhead{} & \colhead{J$1120$+$0641$} &
    \colhead{J$2348$--$3054$} & \colhead{J$0109$--$3047$} &
    \colhead{J$0305$--$3150$}} \startdata
\Mbh(\Mgii) $[10^9$ \Msun$]$ & $2.4^{+0.2}_{-0.2}$ & $2.1^{+0.5}_{-0.5}$ & $1.5^{+0.4}_{-0.4}$ & $0.95^{+0.08}_{-0.07}$ \\ 
\Mbh(\Civ) $[10^9$ \Msun$]$ & $1.09^{+0.02}_{-0.04}$ & -- & $0.77^{+0.05}_{-0.1}$ & $1.20^{+0.06}_{-0.05}$ \\ 
$L_{\rm Bol}/L_{\rm Edd}$ & 0.48 & 0.18 & 0.24 & 0.68 \\
$L_{\rm Bol}/L_{\rm Edd}$$_{\,2011}$ & 0.52 & 0.19 & 0.26 & 0.74 \\
{\SiIV/\Civ} & $0.35\pm0.01$ & -- & $0.39\pm0.19$ & $0.52\pm0.02$ \\
{\CIII/\Civ} & $0.73\pm0.01$ & -- & -- & -- \\
\Feii/\Mgii & $2.10^{+0.13}_{-0.02}$ & $2.8^{+0.3}_{-1.0}$ & $1.8^{+2.5}_{-1.8}$ & $3.2^{+0.7}_{-0.7}$ \\
EW$_{\rm CIV}$ $[$\AA$]$ & $26.3\pm0.3$ & -- & $20.6\pm+4.7$ & $27.0\pm0.8$ \\
\enddata
\tablecomments{${L_{\rm Bol}/L_{\rm Edd}}_{\,2011}$ is obtained by using Eq.~(4) of \citet{DeRosa2011}.}
\end{deluxetable*}

\begin{figure*}
\centering
\resizebox{0.5\textwidth}{!}{\includegraphics{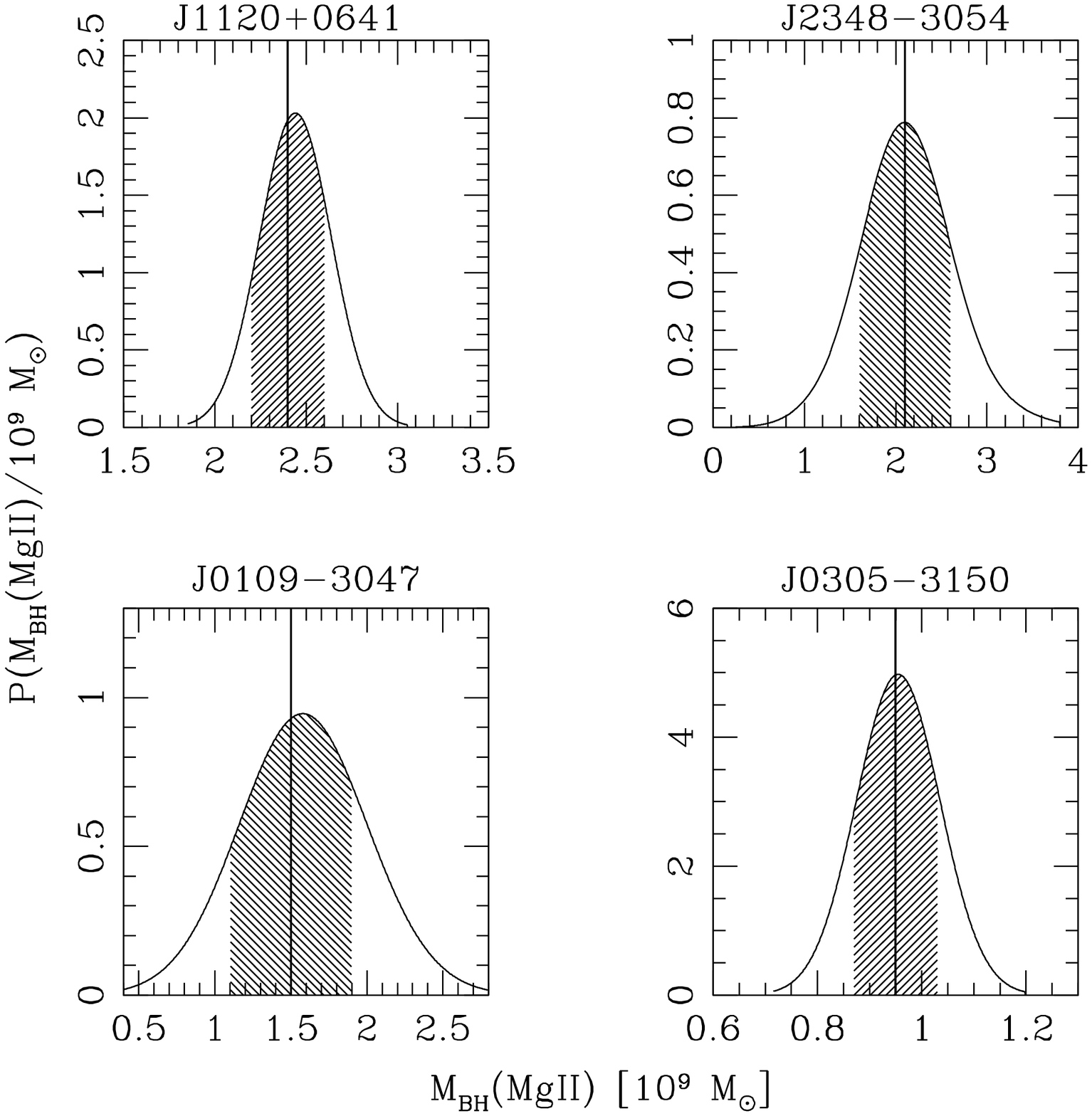}}
\caption{\label{Fig_7} { \Mbh \ estimates from \Mgii \ emission line: marginal probability 
  distributions.  The vertical lines indicate
  the ``best-fit'' estimate, while the shaded areas correspond to the
  1$\sigma$ confidence level (see Section~\ref{sec_procedure} for details).}}
\end{figure*}

In Table~\ref{Tab_derive} we list the \Mbh \ estimates obtained from the \Mgii \
emission line, while in Fig.~\ref{Fig_7} we show the corresponding 
marginal pdfs.
The uncertainties we report do not include the systematic
uncertainties intrinsic to the \Mbh \ estimators.  
{ The \Mbh \ estimate for J$1120$+$0641$ agrees within 1$\sigma$ with the value obtained by
\citet{Mortlock2011} (\Mbh$=2.0^{+1.5}_{-0.7} \times 10^9 \ M_\odot$).}

{ For J$1120$+$0641$, J$0109$--$3047$ and J$0305$--$3150$ we were able to
estimate the \Mbh \ from the \Civ \ emission line (see Table~\ref{Tab_derive}).} 
Even in cases where (a) the \Civ \ emission-line presents extreme blue-shifts
(e.g., for J$1120$+$0641$, $\Delta v \geq 2000$ km s$^{-1}$); (b) the \Civ \
emission-line profile is affected by absorption (e.g., the \Civ \ line of
J$1120$+$0641$ presents an absorption doublet close to the line peak, see
Fig.~\ref{Fig_5} central left panel); and (c) the spectrum is characterized by
low {\it S/N} ratio (e.g., J$0109$--$3047$, see Fig.~\ref{Fig_5} central
panel), we are able to recover the \Mbh \ within a factor of 2.  If we
consider the scatter in the zero point of the two estimators, the \Mbh \
obtained from the \Mgii \ and \Civ \ lines are in agreement within 1$\sigma$.

We further computed the quasar Eddington ratios defined as the ratio between
the AGN bolometric luminosity $L_{\rm Bol}$ and the theoretical Eddington
luminosity $L_{\rm Edd}$.  The AGN bolometric luminosity was obtained by
applying the bolometric correction computed by \citet{Shen2008} to the
observed monochromatic luminosity $\lambda L_\lambda(3000 \rm{\AA})$,
\begin{equation}
\label{eq_bol}
L_{\rm Bol}=5.15 \ \lambda L_\lambda(3000 \rm{\AA}) \ .
\end{equation}
The Eddington luminosity is defined as the maximum luminosity attainable at which the radiation 
pressure acting on the gas counterbalances the gravitational attraction of the BH,
\begin{equation}
L_{\rm Edd}=1.3 \times 10^{38} \left(\frac{M_{\rm BH}}{M_\odot}\right) \rm{erg
  \ s^{-1}} \ .
\end{equation}
{ We computed $L_{\rm Edd}$ from \Mbh(\Mgii).} Since different \Mbh \
estimators can lead to significant differences in the \Mbh \ estimates, and we
aim to compare the Eddington ratios of the $z>6.5$ quasars to the Eddington
ratio distribution of the $4.0<z<6.5$ quasars obtained by \citet{DeRosa2011},
we also re-computed \Mbh$(\mbox{Mg\,\sc{ii}})$ using the same \Mbh \ estimator
as \citet{DeRosa2011}. In Table~\ref{Tab_derive} we list $L_{\rm Bol}/L_{\rm
  Edd}$, and the comparison values obtained by estimating \Mbh \ through
Eq.~(4) of \citet{DeRosa2011}. { The $L_{\rm Bol}/L_{\rm Edd}$ values obtained
through the two estimators are in excellent agreement. The $z>6.5$ quasars are
characterized by an average Eddington ratio of $\langle{\rm log} (L_{\rm Bol}/L_{\rm
  Edd})\rangle=-0.4\pm0.2$, which is in agreement within 1$\sigma$ with the average Eddington 
ratio obtained for the $4.0<z<6.5$ sample: $\langle{\rm log} (L_{\rm Bol}/L_{\rm Edd})\rangle=-0.3\pm0.3$}.

\Mbh \ estimates and relative Eddington ratios of high redshift quasars can be
used to give constraints on the formation processes of super-massive black
holes in the early Universe.  The time needed by a BH seed to grow at a
constant rate from an initial mass $M_0$ to a final mass $M_t$ is equal to
\citep{Shapiro2005}
\begin{equation}
\label{EQ_BH}
t = 0.45 \ {\rm Gyr}  \left(\frac{\epsilon}{1-\epsilon}\right)  \frac{L_{\rm Edd}}{L_{\rm Bol}}  \rm{ln} \left(\frac{M_t}{M_0}\right) {\rm ,}
\end{equation}
where 0.45 Gyr is the is the characteristic accretion timescale (obtained
assuming a mean molecular weight per electron $\mu_{\rm e}=1$), $\epsilon$ is
the radiative efficiency \citep[$\epsilon\sim0.07$,][]{Volonteri2005}, and
$L_{\rm Bol}/L_{\rm Edd}$ is the Eddington ratio. A black hole seed with mass
equal to $M_0=(1$--$6) \times 10^2 \ M_\odot$, which is the hypothesized mass
of a black hole seed originated by a Pop\,III star
\citep[see ][for a review on supermassive black hole (SMBH) formation]{Volonteri2010}, accreting constantly at an Eddington ratio of
$\sim0.4$ (equal to the average Eddington ratio of our $z>6.5$ sample) would
need $\sim$$1.2$--$1.4$ Gyr to grow up to \Mbh$=10^9 \ M_\odot$. However, at
the redshift characteristic of our sources $z=6.5$--$7.1$, the Universe is
only $\sim$0.8 Gyr old while the black holes powering the quasars have
already \Mbh$\gtrsim 10^9 \ M_\odot$.  If we invert Eq.~\ref{EQ_BH}, and we
assume that the black hole seeds grew up to $M_t\sim 10^9 \ M_\odot$ at a
constant Eddington ratio of $\sim0.4$ from $z_0=10,15,\infty$ (accretion time
$t\sim 0.3,0.5,0.8$ Gyr), we find that the BH seed needs to have a mass of
$M_0 \sim 3 \times 10^7,3 \times 10^6,8 \times 10^4 \ M_\odot$ respectively.
Although this is suggesting that highly massive BH ($M_0\gtrsim10^4 \
M_\odot$) seeds are needed at early times ($z\gtrsim20$--$30$) in order to
observe $M_t\sim 10^9 \ M_\odot$ at $z=6.5$--$7.1$, it is extremely difficult
to put robust constraints on the seed masses.  These results are in fact
highly dependent on the adopted Eddington ratio, which, in turn, is a function
of the \Mbh \ estimate (characterized by a systematic uncertainty of $\sim
2$). For example, using the same assumption as above, if ${L_{\rm
    Bol}}/{L_{\rm Edd}}=0.1$ we obtain $M_0 \sim 4 \times 10^8,2 \times 10^8,9
\times 10^7 \ M_\odot$, while for ${L_{\rm Bol}}/{L_{\rm Edd}}=1$ we obtain
$M_0 \sim 1 \times 10^5,4 \times 10^2,1 \times 10^{-2} \ M_\odot$.


\subsection{Emission line properties}


\subsubsection{\rm \SiIV/\Civ \ and \CIII/\Civ \ line ratios}
\label{sec_lines}
As we discussed in Section~\ref{sec_intro}, emission line ratios can be used
to study the BLR chemical enrichment and track its evolution with redshift.
It has been shown that the abundance of nitrogen (N) relative to carbon,
oxygen, and helium (C, O, and He) can be used as a crude marker of the degree
of chemical enrichment of the BLR \citep[e.g.,][]{Hamann2002}.  This is due to
the fact that N is a second generation element, i.e. slowly produced in stars
from previously synthesized C and O. In particular, \citet{Hamann2002} showed
that the most robust abundance probes are N\,{\sc iii}]/O\,{\sc iii}], N\,{\sc
  v}/(C\,{\sc iv}+O\,{\sc iv}), and N\,{\sc v}/He\,{\sc ii}.  Even though the
typical {\it S/N} of our current spectra, together with the severe systematics
affecting the \Lya \ complex, does not allow us to estimate the BLR
metallicity through N\,{\sc v}/(C\,{\sc iv}+O\,{\sc iv}) and N\,{\sc
  v}/He\,{\sc ii} flux ratios, the study of the broad emission line flux
ratios as a function of look-back time does in itself carry significant
information about the BLR chemical enrichment history. \citet{Nagao2006}
analyzed a sample of 5000 quasars from the SDSS Second Data Release: they
built quasar composite spectra in the ranges of $2\leq z\leq4.5$ and $-29.5
\leq M_B \leq -24.5$ mag and measured the emission line ratios in the
composite spectra for each redshift and luminosity bin.  They found that while
there are significant correlations between most of the line ratios and the
quasars luminosities, the flux ratios do not show strong evolution with
redshift. The latter result was further confirmed by \citet{Jiang2007}.

\begin{figure*}
\centering
\resizebox{0.4\textwidth}{!}{\includegraphics{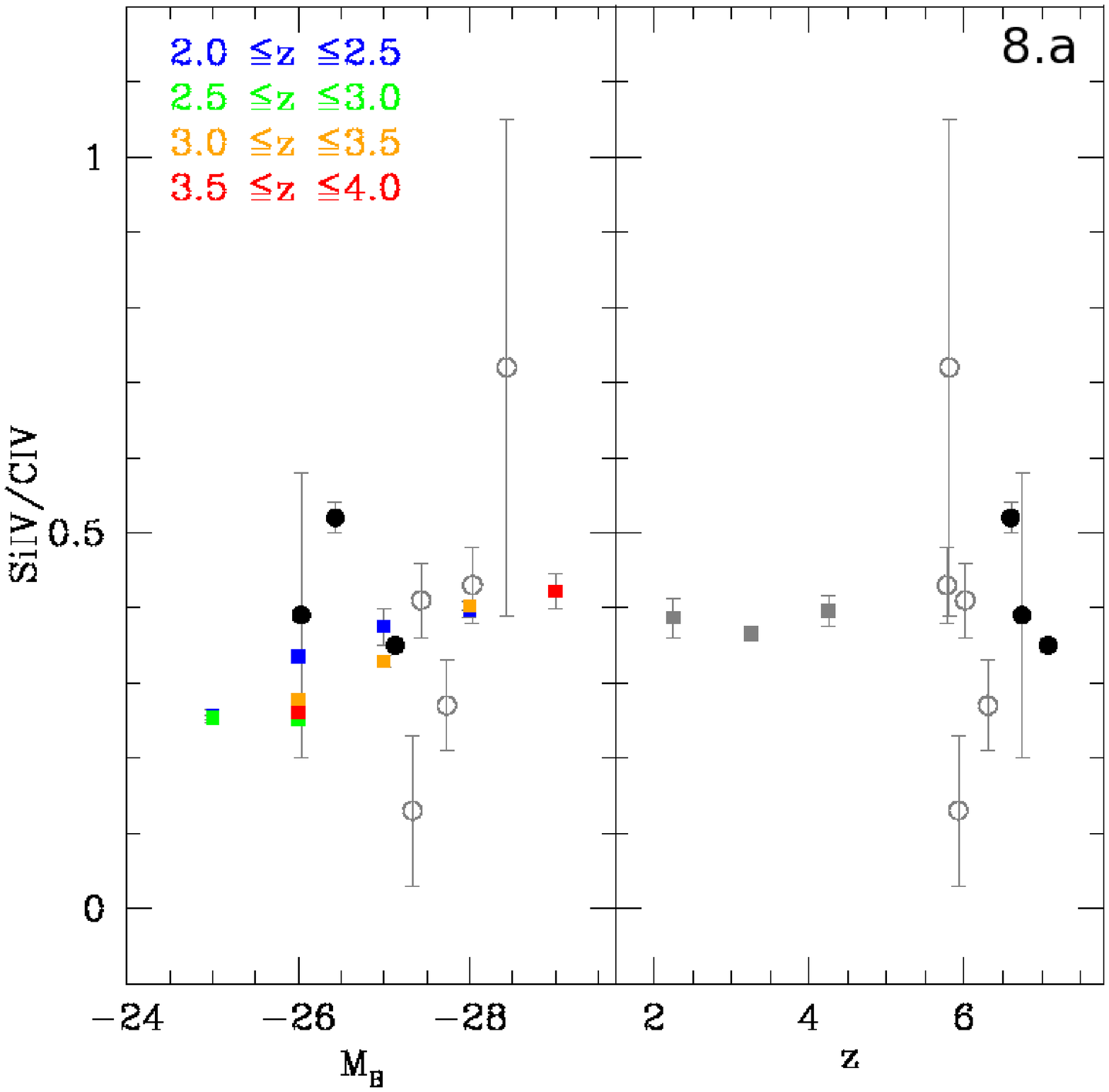}}
\resizebox{0.4\textwidth}{!}{\includegraphics{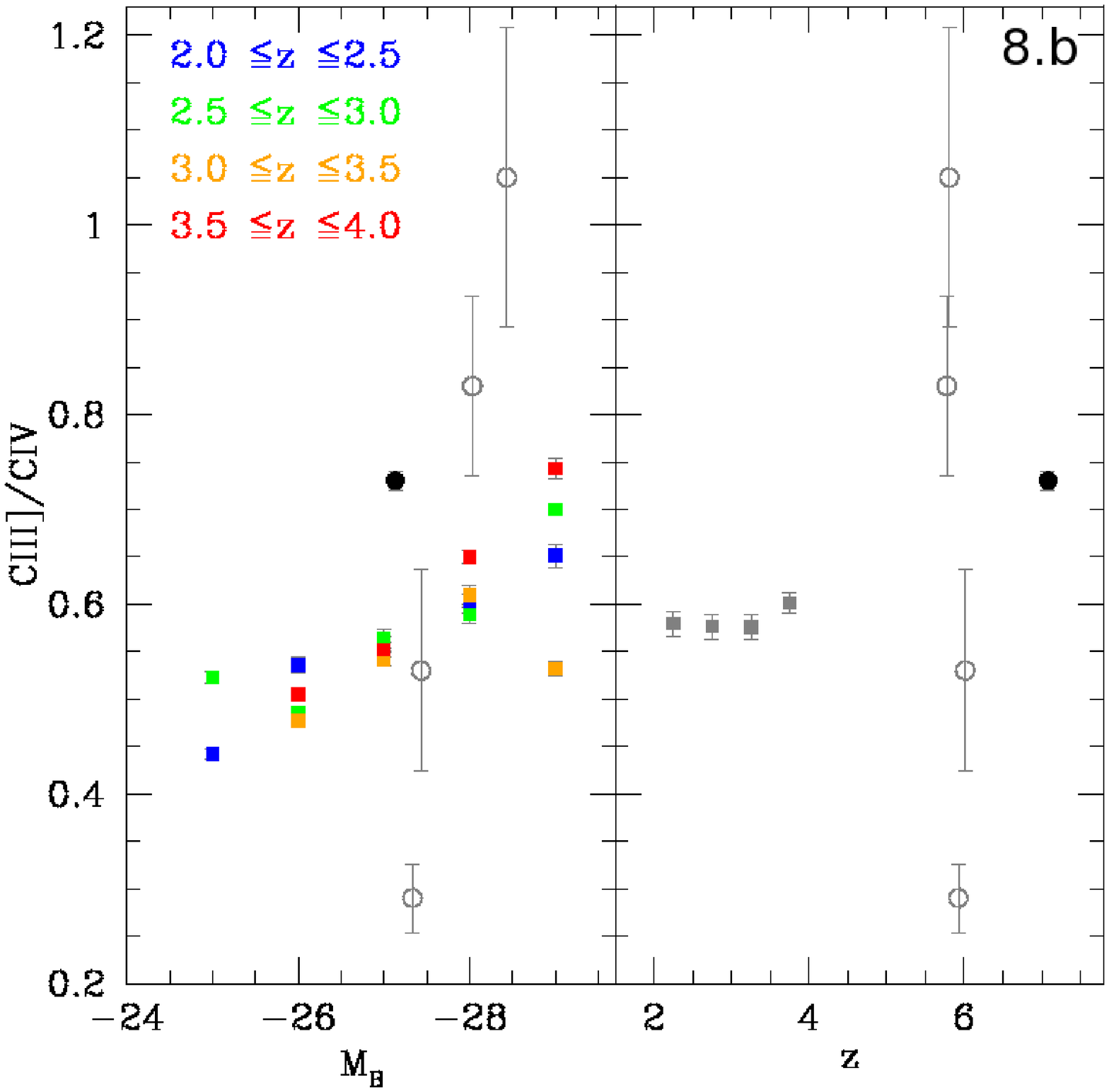}}
\caption{\label{Fig_8} \SiIV/\Civ \ (8.a) and \CIII/\Civ \ (8.b) flux ratios. 
  Left panels: flux ratios as a function of the quasar
  brightness. Colored squares: flux ratios measured in the composite spectra
  of lower redshift quasars, \citet{Nagao2006}; the color coding represents
  the redshift bins (blue: $2.0\leq z \leq 2.5$; green: $2.5\leq z \leq 3.0$;
  orange: $3.0\leq z \leq 3.5$; red: $3.5\leq z \leq 4.0$).  Grey empty
  circles: $z\sim$6 quasars, Jiang et al. (2007). Black filled circles:
  $z>6.5$ sample.  Right panels: flux ratios evolution as a function of
  redshift.  Grey squares: flux ratios measured in the composite spectra of
  quasars in the luminosity range $-28.5 < M_{B} < -26.5$, Nagao et
  al. (2006). Grey circles: $z\sim$6 quasars, Jiang et al. (2007). Black
  filled circles: $z>6.5$ sample.}
\end{figure*}

{ We were able to estimate the \SiIV/\Civ \ flux ratio for J$1120$+$0641$,
J$0109$--$3047$, and J$0305$--$3150$, and the \CIII/\Civ \ flux ratio for
J$1120$+$0641$ (we list the emission line fluxes in Table~\ref{Tab_measure}, while 
flux ratios are listed in Table~\ref{Tab_derive}).} 
In Fig.~\ref{Fig_8} we show the emission line flux
ratios (Fig.~8.a: \SiIV/\Civ; Fig.~8.b: \CIII/\Civ) obtained for the
$z>6.5$ sample as a function of quasar luminosity (left panel) and redshift
(right panel). Together with the results for our sample we plot the flux
ratios of five $z\sim6$ quasars obtained by \citet{Jiang2007}, and the flux
ratios measured in the composite spectra of $2\leq z\leq4.5$ quasars by
\citet{Nagao2006}. In Fig.~\ref{Fig_8} left panels (flux ratios as a function
of luminosity) we plot the results for the lower-redshift sample color-coded
as a function of redshift, while in the right panels (flux ratios as a
function of redshift) we plot only the flux ratios obtained for a sub-sample
of lower-redshift quasars with luminosity comparable to the $z\sim6$ - $7$
objects ($-28.5 \leq M_B \leq -26.5$ mag).  For the high redshift quasars,
$M_B$ was obtained by applying to the respective $M_{1450}$
\citep{Jiang2006,Mortlock2011,Venemans2013} a color conversion factor that was
computed from the SDSS quasar composite spectrum \citep{Vanden2001}.  It is
important to notice that while we are modeling \SiIV+O\,{\sc iv}] and
\CIII+Al\,{\sc iii}+Si\,{\sc iii}] as individual transitions (see
Section~\ref{sec_model_lines}), \citet{Nagao2006} were able to decompose these
emission-line complexes and model their individual components.  At the same
time, \citet{Jiang2007} modeled the Al\,{\sc iii} line in four $z\sim6$
quasars. In cases were a decomposition was performed, we plot the flux
corresponding to the entire emission line-complex, obtained as the sum of the
flux of the individual components.

There is no evidence of evolution of the \SiIV/\Civ \ flux ratio in the
$2.0\leq z \leq 7.1$ redshift range (see Fig.~\ref{Fig_8}.a).  For
the \CIII/\Civ \ line ratio the situation is more controversial. At a given
luminosity, high redshift sources can present flux ratios that are a factor
$\sim2$ different than the ones characteristic of the lower redshift
sample. If we consider the luminosity-matched sub-sample (see
Fig.~\ref{Fig_8}.b, right panel), the \CIII/\Civ \ line ratio does not
show any significant evolution up to $z\sim4$
($<$\CIII/\Civ$>=0.58\pm0.01$). For the $z\sim6$ - $7$ objects instead, we
find significant scatter in the measured line ratios
($0.2\leq$\CIII/\Civ$\leq1.2$). In particular, at $z=7.1$ \CIII/\Civ$=0.57$,
which is $\sim1.3$ times higher than the typical ratio of the
luminosity-matched lower redshift sample. However, given the low number of
sources at $z\geq4$, we cannot draw any definitive conclusion about a possible
dependence of the \CIII/\Civ \ on redshift.


\subsubsection{\rm \Civ \ Equivalent Width}
The equivalent widths (EWs) of high ionization lines anti-correlate with the
underlying AGN continuum luminosity.  The observed degree of anti-correlation
is a function of the line ionization potential \citep[the higher the line ionization
potential, the stronger is the anti-correlation][]{Baldwin1989,Netzer1992}. This relation, also known as
Baldwin effect, was first detected for the \Civ \ emission line by
\citet{Baldwin1977} in a sample of nearby quasars.

\begin{figure*}
  \centering
  \resizebox{0.5\textwidth}{!}{\includegraphics{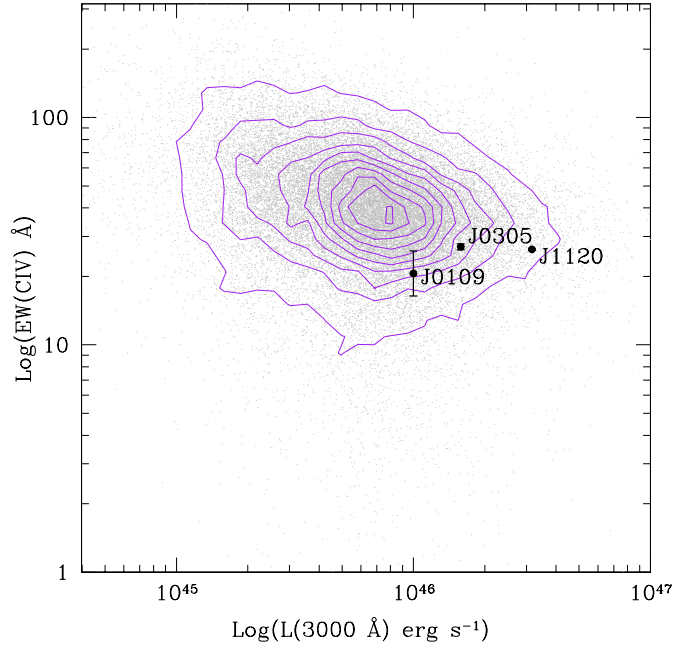}}
  \caption{\label{Fig_9} \Civ \ EW as a function of the AGN continuum
    luminosity. Black point: the $z>6.5$ quasars. Grey dots:
    $1.5 \leq z \leq 2.25$ quasars from the SDSS Data Release 7,
    \citet{Shen2011}. The $z>6.5$ quasars are in agreement with the trend observed for the lower redshift sample.}
\end{figure*}

{ We were able to measure the \Civ \ EW for all the sources in our
sample but J$2348$--$3054$, since it is a BAL quasar.} 
The resulting \Civ \ EWs are listed in Table~\ref{Tab_derive}.
 In Fig.~\ref{Fig_9} we plot the \Civ \ EW as a function of the
AGN continuum luminosity. 
Together with the $z>6.5$ quasars (black points) we are plotting the results
for a sample of 36000 quasars with $1.5 \leq z \leq 2.25$ from the SDSS Data
Release 7 \citep[grey dots, purple contours;][]{Shen2011}. 
{ The $z>6.5$ quasars are in agreement with the trend observed for the lower redshift sample.}

\subsubsection{\rm \Feii/\Mgii \ line ratio}
\label{sec_Fe} 
\begin{figure*}
\centering
\resizebox{0.5\textwidth}{!}{\includegraphics{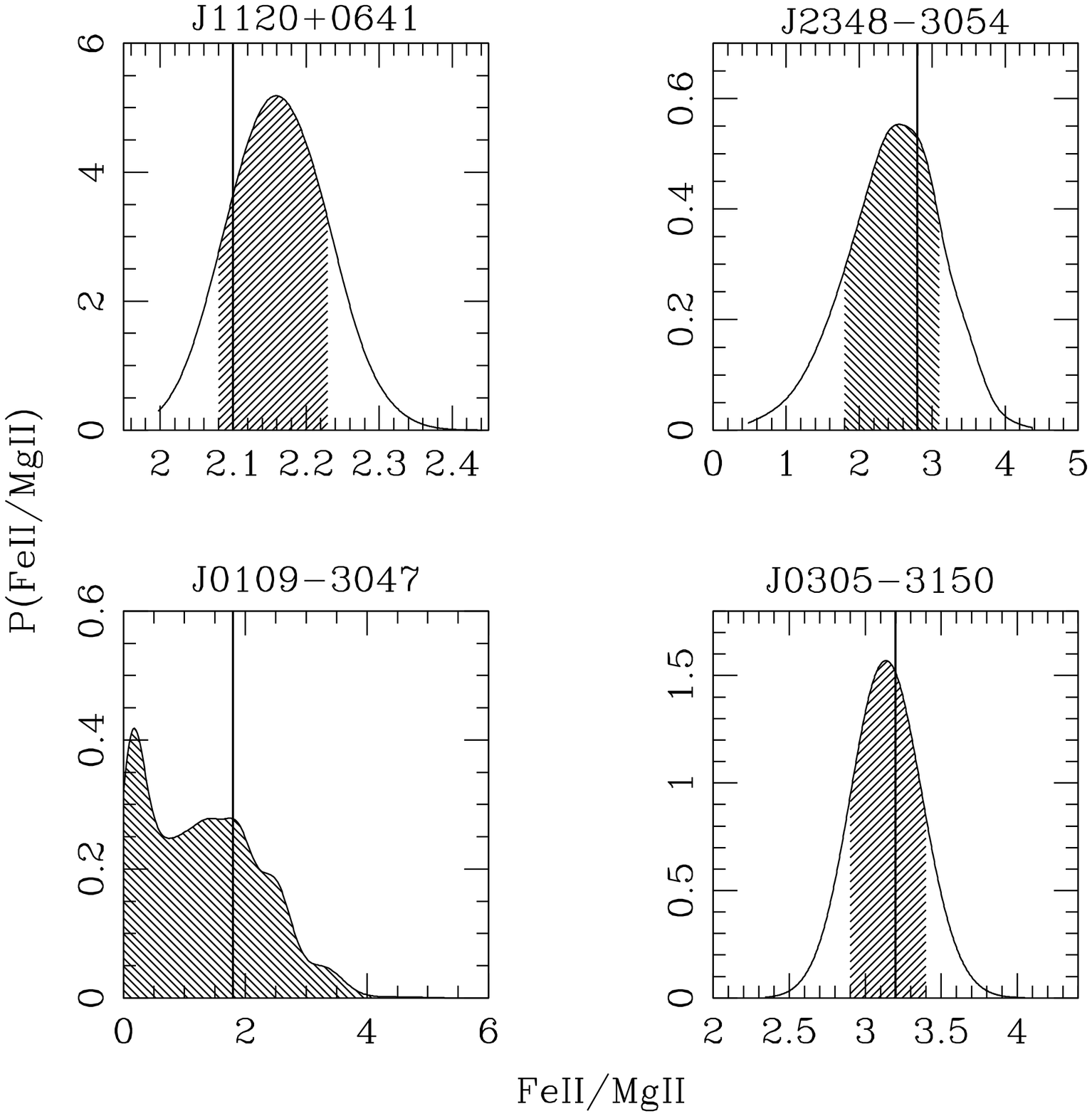}}
\caption{\label{Fig_10} { \Feii/\Mgii \ flux ratios:  marginal probability distributions}.  
  The vertical lines indicate the
  ``best-fit'' estimate, while the shaded areas correspond to the 1$\sigma$
  confidence level for J$1120$+$0641$, J$2348$--$3054$ and J$0305$--$3150$,
  and to the 3$\sigma$ confidence level for J$0109$--$3047$ (see
  Section~\ref{sec_procedure} for details).}
\end{figure*}

As discussed in Section~\ref{sec_intro} the abundance of Fe and Mg is of
particular interest for understanding the chemical evolution of galaxies at
high-z.  We computed the \Feii \ flux by integrating the normalized \Feii \
template over the rest-frame wavelength range $2200 \ \mbox{\AA}
<\lambda_{rest}<3090$ \AA.  { In Table~\ref{Tab_derive}, we list the 
\Feii/\Mgii \ line ratios obtained, while in Fig.~\ref{Fig_10} we show the marginal pdfs.} 
The uncertainties we
report do not include systematic uncertainties.
The resulting pdfs are dominated by the marginal pdf of the normalization of
the \Feii \ template, and are significantly broader for the spectra with 
continuum {\it S/N}$\lesssim$10 (J$2348$--$3054$ and J$0109$--$3047$). At low {\it S/N} in
fact, features such as the \Feii \ complexes, that are significantly fainter
than the bright broad emission lines, become more difficult to detect and,
consequently, the \Feii \ template normalization becomes less constrained. In
particular, for J$0109$--$3047$, the \Feii/\Mgii \ ratio is highly
unconstrained: $0\leq$\Feii/\Mgii$\leq4.3$ with 99.73\% probability
(corresponding to a 3$\sigma$ confidence level).
\begin{figure*}
\centering
\resizebox{0.5\textwidth}{!}{\includegraphics{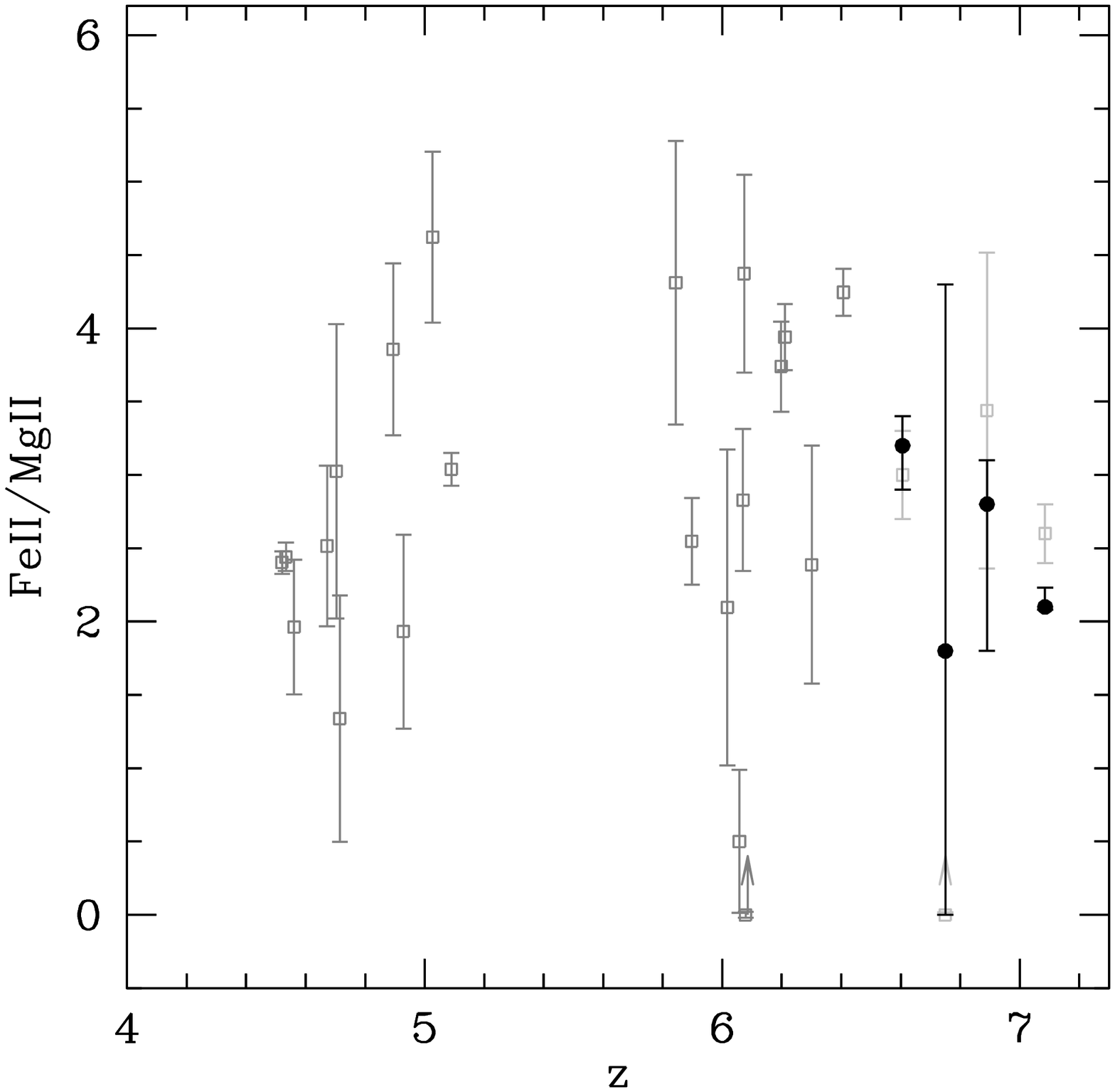}}
\caption{\label{Fig_11} The \Feii/\Mgii \ line ratio as a function of redshift
  for $z>4$.  Black filled circles: $z>6.5$ sample; for J$0109$--$3047$ we are
  reporting the 3$\sigma$ confidence level.  Dark grey empty squares:
  $4.0<z<6.4$ sample, \citet{DeRosa2011}.  Light grey empty squares:
  \Feii/\Mgii \ ratios obtained for the $z>6.5$ sample using the same
  continuum model as \citet{DeRosa2011}.  There is no evidence for evolution
  of the estimated \Feii/\Mgii \ line ratio as a function of cosmic age for
  $4.0<z<7.1$ in the quasar sample.}
\end{figure*}

In Fig.~\ref{Fig_11} we show the evolution of the \Feii/\Mgii \ line ratio as
function of redshift. 
Together with the
$z>6.5$ sample we plot the \Feii/\Mgii \ flux ratios obtained by
\citet{DeRosa2011} for a sample of 22 sources with $4.0<z<6.4$. Even if we are
extending the probed redshift range up to $z\sim7$ (when the age of the
Universe is $\sim0.8$ Gyr), we still do not see any evidence for evolution of
the \Feii/\Mgii \ line ratio as a function of cosmic time.
 
In a previous study \citep{DeRosa2011} we found that the \Feii/\Mgii \ line
ratio measurements are significantly dependent on the adopted modeling
procedure. Therefore, we performed an additional fit of the spectral continuum
after adapting our procedure to the one followed by \citet{DeRosa2011}, where
we considered a sub-class of continuum models with respect to the ones
considered in this work:
\begin{itemize}
\item {we limited the continuum windows to $2000 \leq \lambda_{\rm rest} \leq
    3000$, in order to consider the same broad \Feii \ complexes;}
\item {we considered a single Balmer continuum template with $T_{\rm e}=15000$
    K, and $\tau=1$;}
\item {we fixed the normalization of the Balmer Continuum such that
    $F_{\rm{BC}}(3675 \ {\rm \AA})=0.3 \times F_{\rm{PL}}(3675 \ {\rm \AA})$.}
\end{itemize}
All the resulting \Feii/\Mgii \ line ratios are in agreement within 3$\sigma$
with previous estimates.

To constrain the chemical evolution of the BLR gas in these high redshift
quasars, one would need to connect the measured \Feii/\Mgii \ line ratios with
the corresponding Fe/Mg abundance ratios.  Unfortunately, an accurate
conversion cannot be performed since it has been shown that that the
\Feii/\Mgii \ ratio is not only sensitive to the corresponding Fe and Mg
abundances, but also to the hydrogen density, to the properties of the
radiation field and to the gas micro-turbulence
\citep{Baldwin2004,Verner2003,Verner2004,Bruhweiler2008}.  Therefore,
following \citet{DeRosa2011}, we can conclude that the lack of evolution in
the \Feii/\Mgii \ can be interpreted as an early enrichment of the quasar host
only under the assumption that, in the analyzed sources, the physical
conditions that determine the \Feii \ emission are sufficiently similar.
Under such assumption, the quasar hosts must have undergone a major episode of
Fe enrichment before the cosmic age at which they have been observed
($\sim0.8$ Gyr). On the other hand, if Fe and Mg are produced respectively via
SNe Ia and core collapse supernovae, one would expect Fe to be substantially
produced at least 1 Gyr after the initial burst of star formation
\citep[e.g.,][]{Hamann1993}.  However, the expected enrichment time and our
observations are not in disagreement if we consider that the actual picture is
probably more complex.  Fe could in fact be generated by Pop\,III stars:
extremely metal poor stars with typical masses $M\gtrsim$ 100 \Msun \ that
might be able to produce large amounts of Fe by $z\leq10$ \citep{Heger2002}.
At the same time, stellar nuclear yields are still rather uncertain, and there
are various scenarios in which significant metal production 
can occur at early enough times to obtain a fully enriched BLR at $z\sim7$
 \citep{Matteucci2003,Venkatesan2004}.

\section{Summary}
We have analysed of optical and NIR spectra of the only four $z>6.5$ quasars
known to date: J$1120$+$0641$, discovered in the UKIDSS-LAS survey, and
J$2348$--$3054$, J$0109$--$3047$ and J$0305$--$3150$, recently discovered in
the VISTA-VIKING survey.  We presented new deep VLT/X-Shooter observations for
J$1120$+$0641$.  Together with the new data, we analyzed all the observations
of the $z>6.5$ sources collected by our group using the VLT/X-Shooter
spectrograph and the Magellan/FIRE spectrograph. The collected spectra provide
essentially simultaneous coverage of the $10000$--$24000$ \AA \ wavelength
range. The quality of this data-set is likely the best achievable with the
currently available facilities.

We used the spectra to estimate the masses of the BHs that are powering these
$z>6.5$ quasars and to study their emission-line properties.  The spectra were
modeled using a combination of a power-law continuum, a Balmer continuum, an
\Feii+\Feiii \ template, and a series of emission lines.  We developed a
maximum likelihood procedure for the spectral modeling, which allows a
reliable estimate of how the uncertainties in the continuum modeling propagate
into the estimates of the physical quantities of interest.  The $z>6.5$
quasars are observationally indistinguishable from their counterparts at lower
redshifts.
 
We estimated the \Mbh \ from the \Mgii \ and \Civ \ emission lines using
empirical mass-scaling relations.  The \Mbh \ obtained from the two estimators
agree within 1$\sigma$.  The quasars in our sample host BHs with masses of $\sim
10^9$ \Msun \ that are accreting close to the { Eddington luminosity ($\langle{\rm
  log} (L_{\rm Bol}/L_{\rm Edd})\rangle= -0.4\pm0.2$)}, in agreement with the average 
Eddington ratio obtained for a
$4.0<z<6.5$ sample.  If the measured average Eddington ratio is representative
of the typical quasar accretion rate in the early Universe, highly massive BH
seeds ($M_0\gtrsim10^4 \ M_\odot$) need to be in place at very early times
($z\gtrsim20$ - $30$) in order to be able to observe black holes with
\Mbh$\sim10^9$ \Msun \ at $z=6.5$--$7.1$. At the same time, quasar accretion
episodes characterized by such high rates must be short in time and limited in
number. If the BHs powering our sources had been accreting at an Eddington
ratio of $\sim0.4$ for one additional characteristic accretion time
($t\sim0.45$ Gyr), they would have reached masses of $\sim10^{11}$ \Msun,
which is one order of magnitude larger than the most massive black hole
observed in the local Universe.

We estimated the \SiIV/\Civ \ and \CIII/\Civ \ flux ratios and compared them
with the results obtained from luminosity matched sub-samples at $z\sim6$ and
$2\leq z\leq4.5$. We find no evidence of evolution of these line ratios with
cosmic time.

We calculated fluxes for the \Mgii \ and \Feii \ lines and compared the
measured \Feii/\Mgii \ ratio with the results obtained for a sample of
$4.0<z<6.4$ quasars.  Since the \Feii/\Mgii \ line ratio measurements are
significantly dependent on the adopted modeling procedure, we performed a
consistent analysis of the two samples. We do not detect any redshift
evolution of the \Feii/\Mgii \ ratio for $4.0<z<7.1$.  If we assume that the
\Feii/\Mgii \ line ratio is a reliable proxy of the Fe/Mg abundance ratio,
this indicates that the $z>6.5$ quasar hosts must have undergone a major
episode of Fe enrichment in the first $\sim0.8$ Gyr after the Big Bang.

\begin{acknowledgments}
{ We thank the referee for useful comments that allowed us to improve the quality of the paper.}
GDR and BMP are grateful to the National Science Foundation for support of this work
through grant AST-1008882 to The Ohio State University. BPV acknowledges funding through the ERC grant "Cosmic Dawn"
\end{acknowledgments}

\end{document}